\documentclass[preprint]{aastex}

\slugcomment{ Final revised version: January 10, 2002}

\shorttitle{Meeting the Cool Neighbours I: NLTT sample}

\shortauthors{Reid \& Cruz}

\begin {document}
\title{Meeting the Cool Neighbours I: Nearby stars in the NLTT Catalogue - 
Defining the sample}

\author {I. Neill Reid}
\affil {Space Telescope Science Institute, 3700 San Martin Drive, Baltimore,
MD 21218; \\
Department of Physics and Astronomy, University of Pennsylvania, 209 South 33rd Street, 
Philadelphia, PA 19104
e-mail: inr@stsci.edu}

\author {K. L. Cruz}
\affil {Department of Physics and Astronomy, University of Pennsylvania,  209 South 33rd Street, 
Philadelphia, PA 19104
e-mail: kelle@hep.upenn.edu}

\begin{abstract}
We are currently undertaking a program aimed at identifying previously-unrecognised
late-type dwarfs within 20 parsecs of the Sun. As a first step, we
have cross-referenced Luyten's NLTT proper motion catalogue against the
second incremental release of the 2MASS Point Source Catalogue, and use
optical/infrared colours, derived by combining Luytens's $m_r$ 
estimates with 2MASS data, to identify candidate nearby stars.
This paper describes the definition of a reference
sample of 1245 stars, and presents a compilation of literature
data for over one-third of the sample. Only 274 stars have trigonometric
parallax measurements, but we have used data for nearby stars
with well-determined trigonometric parallaxes to compute 
colour-magnitude relations in the (M$_V$, (V-K)), 
(M$_V$, (V-I)) and (M$_I$, (I-J)) planes, and use those relations
to determine photometric parallaxes for NLTT stars with optical
photometry. Based on the 2MASS JHK$_S$ data alone, we have identified a
further 42 ultracool dwarfs ((J-K$_S > 0.99$) and use (J-K$_S$) colours
to estimate photometric parallaxes. 
Combining these various techniques, we identify 308 stars with formal
distances of less than 20 parsecs, while a further 46 have distance estimates
within 1$\sigma$ of our survey limit.  
Of these 354 stars, 75, including 39 of the ultracool dwarfs. 
are new to nearby star catalogues. Two stars with
both optical and near-infrared photometry are potential additions to the
immediate Solar Neighbourhood,  with formal distance estimates
of less than 10 parsecs. 

\end{abstract}


\section {Introduction}

The scientific bases for completing a thorough survey of the constituents
of the immediate Solar neighbourhood can be grouped under two main
categories: the identification of individual representatives of particular
stellar types for detailed follow-up observation, and the compilation
and analysis of statistical parameters. As individuals, the nearest
stars provide the brightest examples of a particular class, and 
therefore permit the most exhaustive scrutiny of physical characteristics,
and of how those characteristics vary from star to star. Indeed, it is
worth noting that the fact that there are differences in the properties
of individual stars became apparent with the completion of
the first successful determinations of stellar parallax: Henderson's
(1839) analysis of Cape measurements found both components of $\alpha$ Centauri
to be similar in brightness to the Sun, but Bessel's (1836) earlier results
showed that the fainter star in 61 Cygni is $\sim35$ times fainter
than the Sun, while Struve's 1840 observations indicated that Vega
is brighter than the Sun by a similar factor.

From a statistical point of view, 
the scientific justification for compiling a catalogue of the nearest stars 
is summarised succinctly by Kuiper (1942). Besides probing the details of 
stellar evolution through their distribution in
the Hertzsprung-Russell diagram, the nearest stars provide the basis for 
the determination of the stellar
luminosity function, the mass-luminosity relation, the stellar contribution 
to the local mass density, 
the velocity distribution and the stellar multiplicity statistics (including
the frequency of occurence of planetary systems). 
Supplementing the photometric and astrometric
data with chemical abundance determinations, the nearby stars can be used to
map the metallicity distribution of the (local) Galactic disk. Finally, 
with the addition of age estimates, these stars can
probe the local star formation history, and the variation of stellar 
kinematics (and other parameters) as a function of time. 

Success in pursuing Kuiper's research agenda rests on the availability of a
well-defined, representative sample of the local stellar populations. At the
time, no such dataset existed - the most complete sample of nearby stars was 
van de Kamp's (1940) catalogue of 34 systems within 5 parsecs of the Sun. 
Considerable advances have been made in the succeeding three score years, notably through
the efforts of Gliese (1957, 1969), later in collaboration with Jahrei{\ss} (Gliese \&
Jahrei{\ss}, 1979; Jahrei{\ss} \& Gliese, 1991), in compiling results from follow-up observations
of nearby-star candidates identified from a variety of sources. The most recent 
catalogue, the preliminary version of the Third Catalogue of Nearby Stars 
(Jahrei{\ss} \& Gliese, 1991, hereinafter pCNS3), lists over
3800 stars with nominal distances of less than 25 parsecs, although quantitative
spectroscopy (Reid {\sl et al.} - PMSU1, 1995; Hawley {\sl et al.}, 1996 - PMSU2) 
and astrometry show that many stars lie beyond that distance limit\footnote
{Updated measurements for the pCNS3 stars, together with observations of additional, post-Hipparcos nearby star
candidates, are included in the CNS website at {\sl http://www.ari.uni-heidelberg.de/aricns/}.
The PMSU data are available at {\sl http://dept.physics.upenn.edu/~inr/pmsu.html}. }. 
The Hipparcos mission (ESA, 1997) has solidified the local sample of solar-type stars, but
provides data for only a limited subset of stars fainter than 9th magnitude (M$_V = 7.5$, or
spectral type M0, at 20 parsecs). Thus, while the number of known  
nearby systems has increased by two orders of magnitude, the  current M dwarf census
becomes significantly incomplete at distances beyond 10 parsecs. Estimates of
the level of incompleteness vary, ranging from 30-50\% for early and mid-type M dwarfs
to over 75\% at the latest spectral types (PMSU1; Henry, 1998).

The NASA/NSF NStars initiative was designed, at least partly, with the aim of remedying this
notable defect in our knowledge. Working under these auspices, we are undertaking a 
wide-ranging project which aims to use data from the 2-Micron All Sky Survey (2MASS), 
in combination with other large-scale surveys and databases, to 
identify previously-unrecognised late-type dwarfs within 
the immediate Solar Neighbourhood. The near-infrared coverage offered by 2MASS is
ideally suited to detecting and classifying nearby cool dwarfs; indeed, 2MASS 
(Skrutskie {\sl et al.}, 1997) and the
companion DEep Near-Infrared Survey (DENIS - Epchtein {\sl et al.}, 1994), are responsible
for discovering the overwhelming majority of the ultracool low-mass stars and brown
dwarfs which have been used to define the new spectral classes L (Kirkpatrick {\sl et al.},
1999; Mart\'in {\sl et al.}, 1999) and T (Burgasser {\sl et al.}, 2001; Geballe
{\sl et al.}, 2001). 

The prime goal of our project is the identification of all M and L dwarfs within 20
parsecs of the Sun. The near-infrared colours provided by 2MASS are sufficient to
identify late-type M and L dwarfs, but are essentially degenerate for mid-K to M7 dwarfs.
Thus, achieving our goal demands that we employ a variety of techniques,
combining a range of observational strategies. Future papers in this series will discuss the
application of purely photometric selection effects (Cruz {\sl et al.}, in preparation), but
first we concentrate on a variation on a more traditional theme - 2MASS photometry of
stars in the New Luyten Two-Tenths (NLTT) catalogue (Luyten, 1980). Section 2 outlines the relevant
characteristics of the NLTT survey. Section 3 describes the selection
criteria we have used to identify nearby-star candidates, combining the NLTT data with the 
photometry from the second incremental release of the 2MASS point-source catalogue. Section
4 describes the calibration of photometric parallax; section 
5 summarises the data available in the literature for a subset of those sources,
and identifies stars likely to lie within our distance limit of 20 parsecs; section
6 summarises the results. 

\section {The NLTT catalogue and nearby stars}

Proper motion has a well proven track record as a means of identifying nearby stars.
As members of a rotationally-supported, low velocity dispersion system, most\footnote{
but not all - see Reid, Sahu \& Hawley, 2001.} disk dwarfs
have heliocentric space motions of less than 50 km s$^{-1}$. Thus, the
majority of high proper-motion
stars are members of the immediate Solar Neighbourhood - the remainder are high-velocity
members of the Galactic halo. Proper motion determination is also straightforward; measurements
can be made for all stars in a particular region of the sky using wide-field images 
taken at only two epochs. 

The most extensive proper motion catalogues currently available are due to Willem Luyten,
based primarily on his work with the 48-inch Palomar Oschin Schmidt. Attention has mainly
centred on the Luyten Half-Second (LHS) catalogue (Luyten, 1979), which 
includes 3601 stars with  $\mu \ge 0.5$ \arcsec yr$^{-1}$
(and data for a further 869 stars with lower proper motions). This partly reflects
the substantial annual motions of those stars, indicative of either 
close proximity or high space motion, sometimes
both, but also partly reflects the fact that those stars are relatively easy to identify.
Luyten \& Albers (1979) produced the LHS Atlas, which includes finding charts for all of the
fainter LHS stars. The NLTT catalogue, including 58845 stars with $\mu \ge 0.18$
\arcsec yr$^{-1}$, lacks a comparable identification aid. While
the majority of NLTT stars have positions accurate to a few arcseconds, errors exceeding
15 arcseconds are not uncommon.
In searching for the latter targets, astronomers have been known to 
resort to techniques such as using blue and red filters on telescope acquisition systems 
as blink comparators, picking out the reddest (or, for white dwarf candidates,  bluest) 
star in the field. Such methods are far
from efficient, and tend to discourage detailed follow-up observations of extensive target lists
at larger telescopes. 

Luyten's proper motion surveys also offer the disadvantage of low accuracy photometry
(sometime based on by-eye estimates), ill-defined completeness limits and non-uniform
sky coverage. The regions of the celestial sphere accessible from the northern hemisphere
were surveyed in the early 1960s using the Palomar Schmidt, with the original Palomar
Sky Survey (POSSI: Minkowski \& Abell, 1963) 
providing first epoch data. The plates provide both red ($m_r$) and 
blue (m$_{pg}$) magnitude estimates, accurate to $\pm0.5$ mag. and with m$_{pg} \sim B$,
$m_r \sim R_K + 0.8$ (Gliese \& Jahrei{\ss}, 1980; Dawson, 1986).
The faintest stars catalogued have  $m_r \sim 19$ and m$_{pg} \sim 20.5$. 

South of $\delta = -33^o$, both the LHS
and NLTT catalogues are derived primarily from the Bruce Proper Motion survey, which is
based on photographic plates taken with the Harvard 24-inch Bruce refractor. The first
epoch southern hemisphere plates were taken between 1896 and 1910, when the telescope
was located in Arequipa, Peru; Luyten obtained second epoch plates between 1927 and 1929, 
when both he and the telescope were stationed at Harvard's Bloemfontein Observatory in South
Africa. Although the Bruce survey extends to a proper motion limit of 0.1 \arcsec yr$^{-1}$, it
provides only blue-band photographic photometry, and includes
few stars fainter than m$_{pg} \sim 15.5$. 

The absence of deep photographic material at southern declinations is an obvious
limitation in searching for low luminosity dwarfs. However, even
the Palomar data provide far from uniform coverage. The high proper motion stars in the
NLTT are drawn from a relatively small volume, centred on the Sun, so we expect a
uniform distribution over the celestial sphere. Figure 1 plots the 
($\alpha$, $\delta$) distribution of NLTT dwarfs for three magnitude ranges:
$11 < m_{pg} < 14$; $14 < m_{pg} < 15.5$; and $15.5 < m_{pg} < 16$.  Two features are
evident: first, the transition from
Palomar Schmidt data to the Bruce survey, obvious at the faintest magnitudes, but also
discernible at intermediate magnitudes; and, second, the Milky Way.  
The high star density close to the Plane leads to confusion (overlapping images) at
magnitudes well above the POSS I plate limits, and to difficulties in correctly associating
first and second epoch images of
moving objects. It is clear from Figure 1 that at low latitudes, with the exception of a few
regions (such as the Perseus-Auriga region,  $\alpha \sim 5$ hours, $40^o < \delta < 50^o$),
the NLTT catalogue has effectively the same limiting magnitude as the Bruce
survey. 

The number-magnitude distribution of NLTT stars at higher Galactic latitudes is
illustrated in Figure 2, where we also show the distribution of LHS stars in the same
regions. As discussed by Flynn {\sl et al.} (2001), if the kinematics of a stellar population are
invariant over the sampling volume, then the number of stars in a proper-motion limited survey
varies with $\mu_{lim}^{-3}$ (since the distance limit, d$_{lim}$, is inversely
proportional to $\mu_{lim}$). The characteristic distance of a proper motion star also
scales inversely with $\mu_{lim}$, so the typical distance modulus for the 
catalogue scales as  $\mu_{lim}^{-2}$.
Thus, if we compare the number-magnitude distributions of two unbiased proper-motion surveys, 
S$_1$ and S$_2$,
with proper motion limits of $\mu_1$ and $\mu_2$, the sampling volumes scale as
\begin{displaymath}
{\rm {Vol_2 \over Vol_1}} \ = \ f_v \ = \ ({\mu_1 \over \mu_2})^3
\end{displaymath}
 and the relative distance modulus is 
\begin{displaymath}
(m-M)_2 - (m-M)_1 \ = \delta (m-M) \ = \ 5 \log{\mu_1 \over \mu_2}
\end{displaymath}
We need to allow for the change in average distance modulus to ensure 
we are matching stars of similar absolute magnitude. 
Thus, in comparing the number counts, we expect
\begin{displaymath}
N_2 (m) \ = \ f_v \times N_1 (m - \delta(m-M))
\end{displaymath}
In the specific case of the LHS and NLTT surveys, $\mu_1=0.5$ \arcsec yr$^{-1}$
and  $\mu_2=0.18$ \arcsec yr$^{-1}$, so if there are no other selection biases, we
expect
\begin{displaymath}
N_{NLTT}(m) \ \approx \ 21 \times N_{LHS}(m-2.2)
\end{displaymath}
Dawson (1986) estimates that the LHS survey is complete at the 90\% level for $m_r < 18$ 
and $|b| > 10^o$. The LHS therefore provides a reference to $\sim20$th magnitude 
for the NLTT catalogue. 
As Figure 2c shows, scaling the number counts from the two surveys gives a ratio close to
the predicted value for $m_r$(NLTT) brighter than $\sim16$th magnitude, with the
ratio dropping by $\sim20\%$ between 16th and 18th magnitude. This suggests that the NLTT
may be complete only at the 75\% level at the latter magnitudes. 

Despite these limitations, the NLTT catalogue remains a powerful resource for
searching for new candidate stars within 20 parsecs. A proper motion limit of 0.18 
\arcsec yr$^{-1}$ corresponds to a transverse velocity of $\sim 17$ km s$^{-1}$ at 20 parsecs;
simple Monte Carlo simulations based on standard disk kinematics (PMSU2) show that over 85\% 
of nearby stars should exceed this limit.  
Thus, while there is no possibility of using the NLTT to construct a complete census of 
nearby late-type dwarfs, detailed follow-up observations can produce
substantial additions to the number of early- and mid-type M dwarfs known to lie
within 20 parsecs of the Sun. The 2MASS database makes those follow-up observations possible.

\section {The NLTT and 2MASS}

\subsection { Matching the NLTT catalogue against the 2MASS database }

In the near future, 2MASS will provide broadband J, H and K$_s$ photometry for sources
over the full celestial sphere. The J and H passbands match the standard Johnson system, while
the K$_s$ passband, truncated at long wavelengths to avoid terrestial H$_2$O absorption, is 
described and calibrated by Persson {\sl et al.} (1998). The effective
wavelength of the K$_S$ filter is 2.15$\mu$m, as opposed to 2.19$\mu$m for the standard
system, but Carpenter's (2001) analysis reveals only minor differences with respect to
standard systems. In particular, Carpenter finds
\begin{displaymath}
K_S ({\rm 2MASS}) \ = \ K_{CIT} \ - \ 0.024, \qquad 0 < (J-K_S) < 2.9
\end{displaymath}
and
\begin{displaymath}
K_S ({\rm 2MASS}) \ = \ K_{UKIRT} \ + \ 0.004 (J-K_S) \ + 0.002,  \qquad -0.2  < (J-K_S) < 3.8
\end{displaymath}
These differences are negligible compared with other sources of uncertainty
in the present analysis, and we adopt the convention K$_S$=K in this series of papers. 
  
The 2MASS catalogue includes
sources which have a signal-to-noise ratio exceeding 7, corresponding to typical limiting
magnitudes of J$\sim 16.1$, H$\sim 15.2$ and K$_s \sim 14.9$ in uncrowded fields. 
M dwarfs within 20 parsecs have near-infrared
magnitudes significantly brighter than these limits - for example, even an M9.5 dwarf,
comparable to BRI0021 or LP 944-20, has M$_K \sim 11.1$, or K$_s \sim 12.6$ at a distance of
20 parsecs. At those magnitudes, the typical photometric uncertainties 
are 0.02-0.04 magnitudes. 

2MASS survey observations were completed in early 2001, but at the time of writing, 
data are available publicly for only 46.5\% of
the sky via the second incremental release. The results described in this paper, and in
subsequent papers in the series, rest on the latter dataset. 
In addition to photometry, the catalogue
provides astrometry for each source, accurate to $< 1\arcsec$; morphogical information,
allowing segregation of extended and point sources; and a number of data quality flags, identifying
artefacts and potentially confused (in the crowding sense) objects. 

Despite the reservations concerning the NLTT astrometry noted in the previous section, 
positional coincidence is the most effective method of cross-referencing the proper motion
catalogue against the 2MASS database. 
We have applied proper motion and precession corrections to
the NLTT data to transform the co-ordinates to epoch 1998.0 (approximating the mean epoch
of the data in the 2MASS second incremental release)  and equinox J2000.0. We have
cross-referenced this search list against the 2MASS database using the 'GATOR' tool
provided by Infrared Science Archive (IRSA\footnote{http://irsa.ipac.caltech.edu/}),
setting a search radius of 10\arcsec and including only non-extended 
sources. Given the discussion in the previous section, we have also excluded all NLTT dwarfs within
10 degrees of the Galactic Plane. Of the 58845 source in the NLTT catalogue, 23795 (40.4\%) have at
least one 2MASS source within the 10\arcsec search radius; approximately 1400 have two or more
matches, giving a total of 25305 potential near-infrared counterparts to the proper motion
stars. 

This dataset provides the basis for constructing our primary NLTT sample of nearby star
candidates. However, it does not include all of the NLTT stars within the area covered by
the currently-available 2MASS data. We identified those objects by removing the 
matched NLTT stars from the search list, and re-running the database query, but with a
search radius of 60 arcseconds. A total of 4875 additional NLTT stars (8\% of the 
catalogue) have potential 2MASS counterparts at those larger separations\footnote{
A further 853 NLTT stars with $-10^o < b < 10^o$have 2MASS counterparts.}. Figure 3
plots the ($\alpha, \delta$) distribution of the two datasets. It is clear that
the wide-paired NLTT stars (the 4875 stars) are not randomly distributed: there are
obvious concentrations, notably near the North Celestial Pole and near the
Sorth Galactic Pole ($\alpha \sim 1^h, \delta \sim -30^o$). It is likely that these features 
stem from systematic errors in the NLTT positions in those regions.

Figure 3 highlights 
two issues: first, as already discussed, a sizeable subset (20\%?) of the stars in the 
NLTT catalogue have  astrometry of only modest accuracy; 
second, even though 23795 NLTT stars have 2MASS sources within 10\arcsec, there is no
guarantee that those sources include the NLTT star itself. Thus, just as the NLTT catalogue
includes only an incomplete subset of late-type dwarfs with 20 parsecs,  our cross-referencing
against the 2MASS database succeeds in identifying only a subset of the nearby late-type
dwarfs in the NLTT. 
We will discuss the 4875 sources in the NLTT wide-matched sample in a later paper in this series;
for the present, we concentrate on the sample of 23795 NLTT dwarfs with 2MASS
counterparts within 10\arcsec of the predicted J2000 positions. 

\subsection { Colour selection of candidate nearby stars}

Clearly it is unreasonable to attempt detailed follow-up observations of all 25000+
potential NLTT/2MASS pairings. However, we can use Luyten's $m_r$ photometry to pare the
sample to a manageable size. Dawson's (1986) analysis of data for over 2000 LHS stars
confirmed Gliese \& Jahrei{\ss}' (1980) calibration of $m_r$ against standard Kron R$_K$ photometry,
deriving 
\begin{displaymath}
m_r \ = \ R_K \ +  \ 0.80
\end{displaymath}
The (R$_K$-K$_s$) colour spans a long baseline in wavelength, and ranges from $\sim3.0$ at
spectral type M0 to $\sim6.6$ at spectral type M8. Thus, even with uncertainties of
$\pm0.5$ magnitude in $m_r$, the location of a star in the ($m_r$, ($m_r$-K$_s$)) plane
can discriminate between a relatively distant early-type M dwarf and an M6 dwarf in the 
immediate Solar Neighbourhood.

Figure 4 illustrates how we have defined our selection criteria. Since $m_r$ is a
poorly-defined photometric system, with a passband limited to the blue half of more
conventional R passbands, we have not attempted to transform data from standard photometric
systems to define a calibration sequence. Instead, we define the sequence directly, using
photometry listed in the NLTT catalogue for nearby stars with accurate trigonometric
parallax measurements. The near-infrared data for those stars are taken either from
Leggett's (1992) compilation or from the 2MASS survey itself. Figure 4 plots these data, where
the magnitudes are adjusted to match a distance of 20 parsecs. As expected, there is considerable
scatter, so rather than fit a mean relation, we have defined a series of linear relations
which underlie the overall distribution. These provide a set of conservative criteria, 
erring towards including stars lying beyond the 20-parsec limit, rather than excluding nearby
stars with particularly errant photometry. The relations are as follows:
\begin{displaymath}
m_r(lim) \ = \ 2.17 (m_r - K_s) \ + 3.65, \quad  (m_r - K_s) \le 4.3
\end{displaymath}
\begin{displaymath}
m_r(lim) \ = \ 5.25 (m_r - K_s) \ - 9.58, \quad  4.3 < (m_r - K_s) \le 4.7
\end{displaymath}
\begin{displaymath}
m_r(lim) \ = \ 1.48 (m_r - K_s) \ + 8.15, \quad  4.7 < (m_r - K_s) \le 7
\end{displaymath}
We set a lower limit of ($m_r - K_s) = 3.5$, corresponding to (R-K$_s) \sim 2.7$, or spectral
type $\sim$K5, and include all matches with ($m_r-K_s) > 7$. NLTT/2MASS pairings are
eliminated from our candidate list if $m_r > m_r(lim)$. Applying these selection criteria
reduces the NLTT sample by almost 95\%, from 25305 pairings to only 1434 candidates. 

\subsection { NLTT binaries and extreme colours }

Over 2300 stars in the NLTT catalogue are identified in the notes as probable
common proper-motion (cpm) companions of
brighter stars. A substantial fraction of those systems have separations of less than
20\arcsec. Our cross-referencing against the 2MASS database is based only on positional
coincidence, so it is possible for an NLTT binary to produce four pairings: two
correct matches, [NLTT(A)+2MASS(A)] and [NLTT(B)+2MASS(B)]; and two mismatches, 
[NLTT(A)+2MASS(B)] and [NLTT(B)+2MASS(A)]. Of the two mismatches, the latter is more
important for present purposes, since it pairs the fainter optical source against the
brighter infrared source, giving the reddest possible ($m_r-K_s$) colour. Those sources
are most likely be included in our list of nearby-star canddiates.

We dealt with this possible source of contamination through visual inspection (via IRSA) of
the 2MASS images of the cpm companions included in our candidate list. Since
Luyten's notes give the position angle for each system, it is straightforward to
determine whether the 2MASS position corresponds to the correct component. Based on that
comparison, we have eliminated a further 161 pairings, reducing our primary NLTT sample to
1273 candidates and eliminating many of the apparently reddest stars in the sample (Figure 5).

\subsection {Near-infrared colours}

Finally, we have examined the photometric properties of 2MASS sources to check their consistency
with both the ($m_r-K_s$) colours and known properties of late-type dwarfs. Figure 6 plots
the (($m_r-K_s$), (J-K$_s$)) and ((J-H), (H-K$_s$)) two-colour diagrams for the 1273 
NLTT/2MASS pairings which survive as nearby-star candidates. The overwhelming majority have
colours consistent with those expected for M dwarf stars, but there is a small number
of outliers. In particular, 20 sources have near-infrared colours more consistent with either
earlier-type (G, K) main sequence stars or red giants, while a further eight 
have non-stellar JHK colours. 
Figure 6 shows that most of the outliers in the JHK plane (where we have more accurate
photometry) are also discrepant in the optical/near-infrared two-colour diagram; in particular,
the 2MASS sources with early-type near-infrared colours have faint NLTT counterparts, and 
correspondingly red ($m_r-K_s$) colours. Visual inspection confirms that both these objects
and the candidates with red-giant JHK colours are mismatches, and we have eliminated them
from the sample. 

The unusual colours of the remaining outliers can be attributed to an error in one band
of the 2MASS photometry, in some cases probably due to confusion. For completeness, Table 1
lists relevant data for these objects. All are known nearby stars, and at least four
lie within 20 parsecs of the Sun.

\subsection {Summary: NLTT Sample 1 }

With the elimination of mismatches and stars with unreliable photometry, our primary sample
of NLTT nearby star candidates includes 1245 sources. We will refer to these stars as
NLTT Sample 1. 
Figure 7 plots the number-magnitude distribution for the sample, while Figure 8 plots
the distribution on the celestial sphere. 
A relatively small proportion of the sample have faint magnitudes,
with most stars lying between 11th and 15th magnitude and over half brighter than $m_r = 14$. 
More detailed follow-up observations of the latter stars can be obtained in a relatively
straightforward manner using small telescopes. Indeed, such data are already in hand for a
significant fraction of the sample, and these data are discussed in the final sections of 
this paper. Paper II in this series presents BVRI photometry for 180 of the brighter
southern stars in the sample (Reid, Kilkenny \& Cruz, 2001), while Paper III (Cruz \& Reid, 2001)
discusses low-resolution spectroscopy of seventy of the fainter NLTT stars.

\section { Photometric parallax calibration}

The prime goal of our NStars project is identifying stars within 20 parsecs of the Sun. Given the
accuracy possible in current astrometric work (better than 1 milliarcsecond), trigonometric 
parallax measurements offer the most reliable distance estimates. However, acquiring the
necessary astrometric observations remains a time consuming process. Photometric parallaxes, 
derived by estimating the absolute magnitude based on measurement of appropriate colours,
are much simpler to obtain. The main disadvantage is that, since absolute magnitude is calibrated
based on a mean relation, the pgotometric method takes no account of intrinsic scatter in the
HR diagram, due, for example, to abundance variations or unrecognised binarity. Moreover, a
mean relation can smooth over abrupt changes in slope in the main-sequence, leading to 
systematic under- or over-estimates of absolute magnitude in a particular colour range.
Nonetheless, if one bears those caveats in mind, photometric parallax estimates can be
used to further refine the list of nearby-star candidates. 

\subsection {Calibrating the main sequence for nearby stars}

We have chosen three colour indices for calibration purposes: (V-K) is the longest baseline
colour index available for most stars in the sample; (V-I), where I is on the Cousins system, 
is widely used as an optical distance indicator; and (I-J) was identified as the cleanest
optical/near-infrared colour index by Leggett {\sl et al.} (1996). We have calibrated 
the mean relations using data from three main sources: Leggett's (1992) compilation of
UBVRIJHK photometry of nearby K and M dwarfs; a combination of Bessell's (1990) BVRI data 
and 2MASS JHK$_S$ photometry for stars from 
the second Catalogue of Nearby Stars (Gliese, 1969, and Gliese \& Jahrei{\ss}, 1979; hereinafter, CNS2); 
and Dahn {\sl et al's}
(2000) optical and near-infrared photometry of ultracool M and L dwarfs. All of the stars
have trigonometric parallax measurements, derived from either Hipparcos (ESA, 1997) or
USNO observations (Monet {\sl et al.}, 1992 and references therein), accurate to better 
than 10\%; in most cases, 
the accuracy exceeds 5\%, rendering statistical Lutz-Kelker corrections of negligible
proportion. Finally, all known binaries and halo subdwarfs (e.g. Gl 191, Kapteyn's star) have been 
excluded from the sample, together with a few additional stars which lie significantly
above or below the main body of the data. These calibrators should therefore provide a reliable
estimate of the mean location of the main sequence in the local Galactic disk. 

Figure 9 plots the colour-magnitude distribution of main-sequence stars in the (M$_V$, (V-K))
plane. We have derived a mean relation by fitting a sixth order polynomial,
\begin{eqnarray*}
M_V & = & -30.36 + 44.34 (V-K) - 21.84 (V-K)^2 + 5.600 (V-K)^3 - 0.7543 (V-K)^4  \\
 & & \mbox{} + 0.05105 (V-K)^5 - 0.001370 (V-K)^6, \\
 & & \qquad 10 (V-K) > 2.5, \ \sigma = 0.412 {\rm \ mag., \ 198 \ stars}
\end{eqnarray*}
Note the preponderance of datapoints below the mean relation in the colour range $5 < (V-K) < 6$.

Our adopted (M$_V$, (V-I)) calibration is shown in Figure 10. We match the observations using
a composite relation, combining the following three polynomials:
\begin{eqnarray*}
M_V & = & -4.415 + 27.62 (V-I) - 28.45 (V-I)^2 + 14.63 (V-I)^3 - 2.967 (V-I)^4 \\ 
& & \mbox{} - 0.02758 (V-I)^5 + 0.05848 (V-I)^6, \qquad 1.0 \le (V-I) < 2.8, \ \sigma = 0.40 {\rm \ mag., \ 175 \ stars}
\end{eqnarray*}
\begin{displaymath}
M_V \ = \ 12.20 (V-I) - 21.96,  \qquad 2.8 \le (V-I) < 2.9
\end{displaymath}
\begin{eqnarray*}
M_V & = & 5.923 + 2.249 (V-I) + 0.171 (V-I)^2 - 0.01886 (V-I)^3,  \\
& &  2.9 \le (V-I) < 4.5, \ \sigma = 0.22 {\rm \ mag., \ 29 \ stars}
\end{eqnarray*}
As discussed in previous papers (PMSU2; Reid \& Gizis, 1997), this 
tripartite approach
is required by the noticeable steepening of the main sequence at (V-I)$\sim 2.85$.

Finally, Figure 11 plots the (M$_I$, (I-J)) relation. There is clearly an abrupt change in
slope at (I-J)$\sim1.5$, and we have derived separate mean relations for the brighter and fainter
stars, 
\begin{eqnarray*}
M_I & = & \ 2.879 + 1.635 (I-J) + 5.258 (I-J)^2 - 4.516 (I-J)^3 + 1.632 (I-J)^4
- 0.107 (I-J)^5, \\ & & 0.4 \le (I-J) < 1.45, \ \sigma = 0.42 {\rm \ mag., \ 194 \ stars}
\end{eqnarray*}
\begin{eqnarray*}
M_I & = & 16.491 - 16.499 (I-J) +14.003 (I-J)^2 - 4.717 (I-J)^3+ 0.697 (I-J)^4
- 0.0330 (I-J)^5 , \\ 
& &  1.65 \le (I-J) < 4.0, \ \sigma = 0.31 {\rm \ mag., \ 37 \ stars}
\end{eqnarray*}
The main sequence is essentially vertical in region of overlap, with an almost
even distribution of datapoints over the range ($1.45 < (I-J) < 1.65$, $9.2 < M_I < 11.2$).
Rather than attempt to fit a mean relation, we assign an absolute magnitude 
estimate of $M_I = 10.2\pm0.7$ for NLTT stars falling in this colour range. 

\subsection {Structure in the main sequence}

The disk main sequence does not, unfortunately, present a simple linear relation
in colour-magnitude diagrams - hence the necessity for the
polynomial relations computed in the previous section. 
Before applying those calibrations to derive photometric parallaxes for the NLTT stars,
we briefly consider both the interpretation of the changing slope of the main sequence
evident in Figures 9, 10 and 11, and the implications for our analysis. 

A change in slope of the main sequence in a colour-magnitude diagram generally 
reflects either a
significant change in the opacity distribution within the individual spectral bands sampled
(a local effect), or a significant change in the underlying physical structure (a global
effect). The most striking example of the former is the abrupt change in near-infrared 
(H, K) colours at the transition between spectral types L and T 
due to the onset of CH$_4$ absorption at 1.6 and 2.2 $\mu$m. 
In contrast, most of the changes in slope evident in Figures 9-11 likely 
stem from global effects.

Several notable points of inflection are evident in Figures 9 and 10:
 at M$_V \sim 8.5$ (spectral type M1), where the main-sequence steepens; at
M$_V > 14$ (spectral type M4.5/M5), where the gradient becomes shallower, almost matching the slope
at M$_V < 8$;  and, less pronounced in (V-K) but nonetheless present, at
M$_V \sim 12.5$ (spectral type M3.5/M4), where the the main sequence steepens sharply. 
The `break' in the main-sequence produced by the latter two
points of inflection is evident at near-infrared wavelengths at (I-J)$\sim 1.5$,
while PMSU2 and Reid \& Gizis (1997) have shown that this
feature is also present if one uses TiO bandstrength as a surrogate for colour (effective
temperature). We emphasise that the same stars outline the configuration at all wavelengths:
thus, Gl 15B (M$_V$=13.33, (V-I)=2.82, M$_I$=10.51, (I-J)=1.48, M3.5) is one of the bluest and faintest 
contributors, while Gl 555 (M$_V$=12.36, (V-I)=2.86, M$_I$=9.50, (I-J)=1.59, M4)
lies at the opposite
extreme. The fact that this feature occurs over such a wide range in wavelength, coupled with
the lack of any obvious rapidly-varying spectral features, suggests strongly that this is
a global effect, indicative of a significant change in luminosity over a small range in
colour (effective temperature). In contrast, the steepening in the (M$_V$, (V-I))
distribution at M$_V > 18$, behaviour which is not reflected in (V-K), is probably 
a local effect, 
marking the presence of substantial TiO and metal hydride absorption in the I-band. 

Several theoretical mechanisms are known to modify the shape of the lower
main-sequence. At masses below $\sim0.1 M_\odot$, degeneracy becomes increasingly
important, leading to the shallower slope at M$_V > 13$ (D'Antona \& Mazzitelli, 1985).
On the higher luminosity side of the break,  
Copeland {\sl et al.} (1970) originally demonstrated that 
H$_2$ formation affects the atmospheric temperature structure in late-K and early-M
dwarfs. At those temperatures the formation region lies in the convection zone, leading
to a shallower adiabatic gradient, a higher luminosity and a higher surface temperature  
for stars below the threshold mass. Copeland {\sl et al.} 
place the onset of this effect at M$_{bol} \sim 7$, broadly consistent with the observed
change of slope at M$_V$=8.5. More recent theoretical calculations by Kroupa, Tout \& 
Gilmore (1990), 
on the other hand, find a lower threshold luminosity, M$_{bol} \sim 9$, or M$_V \sim 11$.

As yet, there is no widely-accepted theoretical explanation for the break in the 
main-sequence at (V-I)=2.8. 
Clemens {\sl et al.} (1998) suggest that the feature may be a result of a relatively abrupt
decrease in radius, possibly correlated either with the onset of full convection, or 
an internal change in the structure of the core. In any event, 
none of the available theoretical models reproduce the observed main sequence
at these luminosities. As an illustration, 
Figures 9, 10 and 11 plot the 5-Gyr isochrone from the solar abundance 
models calculated by Baraffe {\sl et al.} (1998 - BCAH), together with 5-Gyr
isochrones form the more recent DUSTY models (Chabrier {\sl et al.}, 2000). 
The latter include both grain opacities and an improved TiO line list, although
incompleteness in the H$_2$O line list leads to inaccuracies at near-infrared
wavelengths (Chabrier, priv. comm., 2001; see also Reid \& Cruz, 2002, for comparison
against infrared data for late-type dwarfs). 

The BCAH models are a closest to the empirical main sequence in 
the (M$_I$, (I-J)) plane, albeit to some extent smoothing over the break at
M$_I \sim 10.5$. The extremely red colours at low luminosities reflect the absence of
grain opacities in those models; the DUSTY models are clearly a better match to the data. 
At optical wavelengths, the BCAH models show poorer agreement, falling 
below the main sequence at M$_V \sim 10$ and remaining
0.5 to 1  magnitudes fainter than the observations at lower luminosities\footnote{
This mismatch accounts for the remarkably young age of $\sim30$ Myrs. deduced 
for Gl 229A by Leggett {\sl et al.} (2002). Since the BCAH models fall below the empirical
main sequence, the only means of matching the observed luminosity is by reducing the age.}
Again, the DUSTY models are better match the data, reflecting the more
extensive TiO linelists, but these models still miss the M3/M4 break in (M$_V$, (V-I)), 
while the mismatch at near-infrared wavelengths reflects the H$_2$O opacity
deficiencies. 
Bedin {\sl et al.} (2001) point out similar discrepancies between theory and
observation at lower abundances. As the
latter authors emphasise, resolving those discrepancies is important both
in interpreting colour-magnitude diagrams, and in establishing 
reliable theoretical mass-luminosity transformations.

In terms of the present survey, structure in the main sequence has two consequences: first,
systematic miscalibration, if the colour-magnitude relation we adopt
fails to follow the empirical distribution;
second, higher Malmquist bias, and a consequent increased contamination from more distant
stars, at colours where the main-sequence is steepest. Both of these biases are likely
to be most significant near the break at M$_V =12$ to 14 ( $5 < (V-K) < 5.6$, 
$1.45 < (I-J) < 1.65$).
These effects will be taken fully into account in statistical analysis of the nearby star
sample. For present purposes, we simply note the increased uncertainty in photometric parallax
for stars of the appropriate colours. 

\section { Literature data for NLTT Sample 1}

We have used the SIMBAD database to cross-reference the NLTT sample against the
published literature, checking all potential named counterparts within 1 arcminute of the
2MASS position. The latter step is essential since SIMBAD does not include cross-references to all
of the LP names cited in the NLTT, while some stars appear twice (or more)
with different names and slightly different positions. Moreover, a significant number of
stars in the NLTT catalogue have no associated name - a deliberate choice on Luyten's part. 
The overwhelming majority of these stars are actually  
from the Lowell Observatory proper motion survey (Giclas, Burnham \& Thomas, 1971). 
Over 400 stars in the sample as a whole prove to have either photometric
or astrometric observations available in the literature.

\subsection {Photometry and astrometry}

Amongst the 1245 stars in our primary NLTT sample, 648 have at least V-band photometry, 
of which 469 are considered here (the remaining 180 stars will be discussed in Paper II).
Three hundred and forty-two of the 469  are listed in the preliminary version
of the third Nearby Star Catalogue (pCNS3, Gliese \& Jahrei{\ss}, 1991), including a number
of known spectroscopic or small angular-separation binary systems. While the latter
are not photometric outliers, unlike the stars listed in Table 1, photometric parallaxes
will lead to underestimated distances, so we have culled those stars from the sample.
Data for those systems are listed in Table 2.
Table 3 collects published photometry and parallax measurements for the remaining stars.
We list the NLTT designation for each, adding the Giclas numbers ignored by Luyten, 
and give Gl or GJ numbers (as appropriate) as a secondary identification. We have also
cross-referenced the sample against the LHS catalogue. 

All of the optical photometry included in Table 3 is on the Johnson/Cousins BVRI system.
The original RI photometry is taken from sources which use either the Kron or the Kron-Cousins
system, since experience has shown that transforming data for M dwarfs from other
systems can give unreliable results. We have used the relations given by
Bessell \& Weis (1987) to transform between the Kron and Cousins systems. 
The main contributor is Weis, who has obtained optical data for nearly 3000  NLTT
stars, including all m-class stars with $\delta > 0^o$ and
$m_r < 13.5$ (Weis, 1988 and refs within), together with  almost 25\% of the LHS
catalogue (Weis, 1996). Two hundred and sixty of those stars are included in Table 3. 
Other sizeable contributions are from Bessell 
(1990 - BVRI, 46 stars), the Hipparcos catalogue (ESA, 1997 
- BV, 43  stars), the pCNS3 (BV - 28 stars) and  Sandage \& Kowal (1986 - BV, 27 stars). 
We also include photometry by Ryan (1989), Fleming (1998), Patterson {\sl et al.} (1998) and 
Eggen (1987).

Figure 12 superimposes photometry for the NLTT stars on the
 the two-colour ((B-V), (V-K$_S$)) and ((V-I), (V-K$_S$)) diagrams outlined by nearby
main-sequence stars. In most cases, the data are broadly consistent with the 
expected distributons, albeit with significantly more scatter in the ((B-V), (V-K$_S$))
plane. A few stars require special comment:
\begin{itemize}
\item LP 335-13 (HIP 91489): the (B-V) colour listed in the Hipparcos catalogue ((B-V)=0.68) 
is clearly
incompatible with both the observed spectral type (M2) and the absolute magnitude inferred
from the apparent magnitude and parallax (M$_V = 8.71$). Since the V magnitude (10.85)
is consistent with the NLTT photometry ($m_r=11.0$), we adopt that value in computing (V-K$_S$).
\item LP 984-91 (HIP 112312): the V magnitude listed in Table 3 is derived 
from the Hipparcos H$_p$ measurement, adopting the colour correction appropriate to a mid-type M
dwarf. We note that the Hipparcos measurements indicate variability of $\sim0.35$ magnitudes.
\item LP 653-13 (LHS 176): the optical colours listed in Table 3 (from Dawson \& Forbes, 1989)
are inconsistent with the
both the inferred (V-K$_S$) and the JHK$_S$ colours, perhaps due to misidentification.
Further observations are required, and 
the (V-K$_S$) photometric parallax computed here must be regarded as tentative.
\item LP 469-50 is clearly identical with G 3-34. Inspection of POSS I and II images, however,
shows that the position listed in SIMBAD for the
latter star is coincident with a nearby, non-moving  star of similar magnitude, lying $\sim2.5$
arcminutes NW of the proper motion star. It is not clear which star was observed by Sandage \& Kowal,
so the  (V-K$_S$) photometric parallax requires confirmation.
\item +19:5093B: the (B-V) colour derived by Eggen may be affected by the presence of the
nearby 6th magnitude primary star.

\end{itemize}

The trigonometric parallax data are from two main sources: The Hipparcos cataloge 
(ESA, 1997 - 141 stars); and the Fourth edition of the Yale parallax catalogue
(van Altena {\sl et al.} (1995). Our sample includes a number of fainter components
in binary systems which lack direct trigonometric parallax measurements, but where
such data are available for the primary star in the system. 
Of the 469 stars in Tables 2, 3 and 4, 178 lack trigonometric parallax data.

\subsection {Distance estimates }

We have used the absolute magnitude/colour relations defined in Section 4.1 to estimate
distances to each star with photometry in the appropriate passbands. Table 4 lists the results,
expressed as distance moduli, and associated uncertainties, $\epsilon$. We have combined the
available individual measurements, weighted by the uncertainty, to derive the average
photometric parallax, (m-M)$_{ph}$. As discussed in \S 4, photometric distance estimates 
cannot take into account the intrinsic dispersion of the main sequence; combining the individual
estimates therefore provides a more precise estimate of the average absolute magnitude
of a star with the observed colours, rather than a more precise estimate of the distance
to a particular star. Trigonometric parallax measurements offer the best method
of measuring distances to individual stars, and 
our dataset includes a substantial number of stars with accurate
astrometry. None of those stars are amongst the photometric calibrators used to define
the colour-magnitude relations given in \S4.1. We can therefore use these additional
stars to verify the reliability of those relations.

Including stars from Paper II in this series, we have optical photometry for 253 stars with
trigonometric parallaxes measured to a formal accuracy better than 9\%. 
Figure 13 plots the residuals in distance modulus for that, in the sense
\begin{displaymath}
\delta (\pi - {\rm phot}) \ = \ (m-M)_\pi \ - \ (m-M)_{phot}
\end{displaymath}
as a function of absolute visual magnitude. Table 5 lists the mean residual
and the dispersion in residuals for the individual photometric estimates and
for the averaged photometric parallax. The rms dispersion is typically 0.3 to 0.4
magnitudes, rising sharply 
in the M$_V$=13 bin, centred on the main-sequence break discussed in section 4.2, but
there is no evidence for a systematic offset. 

Table 4 also lists the trigonometric distance estimates. Since the measured uncertainties,
$\sigma_\pi$, are symmetric in parallax, the uncertainties in
distance modulus are asymmetric. For present purposes, we adopt
\begin{displaymath}
\epsilon_\pi \ = \ 5 \log{{\pi \over {\pi-\sigma_\pi}} }
\end{displaymath}
and use those values as weights in averaging (m-M)$_\pi$ and (m-M)$_{ph}$. Based on
the above discussion and
the comparison shown in Table 5, we set a lower limit of $\pm0.3$ magnitudes on the
weight associated with (m-M)$_{ph}$ to take into account the intrinsic dispersion
of the main sequence. This ensures that high-accuracy trigonometric measurements
are given due weight, while preserving a self-consistent distance estimation process.
Our final adopted estimate of the distance to each star, d$_f$, and the associated 
uncertainty, $\epsilon_d$ are listed in Table 4. 

The last column of Table 4 identifies which stars are likely to lie within 20 parsecs
of the Sun. Stars with formal distances d$_f \le 20$ parsecs are
identified as probable inhabitants of the immediate Solar Neighbourhood ( Y - 266 stars), while candidates with 
d$_f-\epsilon_d \le 20$ parsecs
are possible members ( ? - 46 stars). One hundred and fifty-seven stars have formal distances 
d$_f-\epsilon_d > 20$ parsecs, and are therefore excluded from our census. Amongst the Solar Neighbourhood
members, 43 have formal distances of less than 10 parsecs (identified as Y* in Table 4). While most 
are well-known, much-studied nearby stars with accurate trigonometric parallax measurements, two
stars are potential additions
\begin{itemize}
\item G 39-29, with a formal distance of $9.6\pm1.3$ parsecs amd M$_K$=7.4; no trigonometric
data.
\item G 180-11, $d_f = 9.3\pm1.2$ parsecs and M$_K$=8.1;  no trigonometric
data.
\end{itemize}
Both are listed in the pCNS3, but with higher distance estimates.
Accurate trigonometric parallax data are required to confirm the photometric
distance estimates. 

\subsection {Late-type dwarfs }

The Solar Neighbourhood census is least complete for stars of low luminosity. Early- and mid-type M
dwarfs have near-infrared colours spanning a very small range in magnitude; in particular, (J-K) is
essentially constant, at (J-K$_S)=0.9\pm0.1$ for spectral types K7 to M6. 
The coolest main-sequence 
stars, ultracool dwarfs with spectral types later than M6, have sufficiently extreme energy
distributions that (J-K$_S$) changes significantly with decreasing temperature. We can
therefore identify the ultracool dwarfs in our NLTT sample, and use the near-infrared colours
to estimate photometric parallax. 
Gizis {\sl et al.}  (2000) have calibrated this relation, deriving
\begin{displaymath}
M_K \ = \ 7.593 \ + \ 2.25 (J-K_S), \qquad \sigma = 0.36 \ {\rm mag.}
\end{displaymath}
valid for spectral types later than M6.5.
We have used this relation to estimate distances to NLTT dwarfs in the current sample with
(J-K$_S) > 0.99$. Tables 6 and 7 present the results.
Table 6 lists nine  dwarfs with previous  spectroscopic observations, including
LHS 2090, an M6.5 dwarf recently identified as lying within the 8-parsec sample (Scholz 
{\sl et al.}, 2001); LP 944-20, the nearest isolated brown dwarf (Tinney, 1998); four dwarfs 
from the ultracool 2MASS sample selected by Gizis {\sl et al.} (1999); 
and an earlier type dwarf, LP 860-46,
which appears coincident with one of the brighter stars in Ardila {\sl et al.'s} (2001) U Sco 
photometric survey. 

Table 7 collects data for a further 42 ultracool dwarfs selected from our current sample based
on the 2MASS photometry. We have estimated distances to these dwarfs using the (M$_K$, (J-K$_S$))
relation given above. While the majority of these stars have no prior observations, nine dwarfs
have optical photometry. Photometric parallaxes derived from the latter data (usually
(V-K$_S$)) indicate larger distances than the (J-K$_S$) calibration. 
Indeed, the optically-based distances for the three brightest Giclas stars are a factor of
four higher than the near-infrared calibration. These stars probably have spectral types earlier than
M6.5, but have near-infrared colours on the red extreme of the (J-K$_S$) distribution. The agreement
between d$_f$ and d$_{J-K}$ is better amongst the fainter (apparent magnitude) stars in Table 7 (which are also
likely to have fainter absolute magnitudes), although the near-infrared colour index still tends to give
lower distances by $\sim30\%$. Nonetheless, all of the dwarfs listed in Tables 6 and 7 have formal distances 
either of less than 20 parsecs, or within 1$\sigma$ of our distance limit. Of the 51 ultracool
dwarfs in Tables 6 and 7, only LP 944-20 has a trigonometric parallax measurement.

\section {Summary}

Our NStars survey aims to identify late type stars and brown dwarfs lying within 20
parsecs of the Sun. In this first paper, we have concentrated on defining an initial sample of 
nearby-star candidates from the NLTT catalogue by combining Luyten's red magnitude estimates with
near-infrared photomety from the 2MASS database. We also describe a number of techniques which
will be used in subsequent papers, both to identify other nearby-star candidates and to
estimate their distances. 

Cross-referencing our initial sample against the literature, we have located optical
photometry for 469 of the 1245 stars. We have also used the near-infrared data provided by
2MASS to identify a further 41 ultracool dwarfs. Most of the stars in the former sample were
already known to lie within the immediate Solar Neighbourhood, and are included in the
preliminary version of the Third Catalogue of Nearby Stars. Our re-analysis provides 
improved distance estimates to many of these objects. Three hundred and fifty-six 
stars listed in Table 3 have formal distances of less than 25 parsecs, the distance limit of
the CNS2 and pCNS3; 45 of those stars have no pCNS3 designation. Our analysis also indicates that 
all 51 dwarfs listed in Tables 6 and 7 (ten stars are included in Table 3) also 
meet the pCNS3 distance limit. 
Two hundred and ninety stars from Table 2 and all of the stars in Tables 6 and 7 meet 
the formal criteria of our own survey, with a more modest distance limit of 
20 parsecs. Thirty-seven of the
former sample, and 40 of the latter, are additions to the 20-parsec nearby-star census.

Future papers in this series will present more detailed observations of the less well-studied
stars discussed in this paper, notably the ultracool dwarfs, and of the remaining 735 
stars in our initial NLTT sample.
In addition, we will apply the techniques outlined here in analysis of the 4875 NLTT dwarfs which
were not included in the parent sample discussed here, but have  potential
2MASS matches within 60\arcsec. 

\acknowledgements 
The NStars research described in this paper was 
supported partially by a grant awarded as part of the NASA Space 
Interferometry Mission Science Program, administered by the Jet Propulsion Laboratory, Pasadena.
This publication makes use of data products from the Two Micron All Sky Survey, which is
a joint project of the University of Massachusetts and the Infrared Processing and Analysis 
Center/California Institute of Technology, funded by the National Aerospeace and Space 
Administration
and the National Science Foundation. We acknowledge use of the NASA/IPAC Infrared Source 
Archive (IRSA), 
which is operated by the Jet Propulsion Laboratory, California Institute of Technology, 
under contract with the  National Aerospeace and Space Administration.
We also acknowledge making extensive use of the SIMBAD database, maintained by Strasbourg 
Observatory,
and of the ADS bibliographic service. Finally, we thank the anonymous referee for useful comments.

\newpage

\begin{table}
\begin{center}
{\bf Table 1 } \\
{Photometric outliers}
\begin{tabular}{rrrrccrccrccr}
\tableline\tableline
NLTT & $\alpha$ (2000) & $\delta$ & $m_r$ & ($m_r - K_S$) & (J-H) & (H-K$_S$) & $\pi$ & M$_K$ \\
\tableline
G 74-34 & 02 \ 36 \ 47.8 & 32 \ 04 \ 20 & 12.6 & 4.33 & 0.72 & 0.03 & $65.2\pm1.5$ & 7.35 \\
GJ 1194B & 15 \ 40 \ 03.7 & 43  \ 29 \ 35 & 13.0 & 4.75 & -0.37 & 1.02 & $74.2\pm4.8$ & 7.60 \\
LP 229-17 & 18 \ 34 \ 36.6 & 40 \ 07 \ 26 & 11.5 & 4.42 & 0.67 & -0.55 & $138\pm40$ &  \\
+46:2654 & 19 \ 16 \ 11.7 & 47 \ 05 \ 13 & 11.2 & 3.54 & 0.68 & -0.32 & $36.2\pm1.4$ & 5.45 \\
+48:3952B & 23 \ 10 \ 21.4 & 49 \ 01 \ 02 & 10.0 & 3.67 & 0.31 & -0.02 & $21.6\pm0.9$ & 3.00\\
G 273-93 & 23 \ 38 \ 08.1 & -16 \ 14 \ 09 & 12.3 & \nodata & 0.60 & \nodata & $62\pm18$ & \\
\tableline\tableline
\end{tabular}
\end{center}
Notes: \\
G 74-34: binary, $\delta$V=0.3 mag. (pCNS3), parallax from van Altena {\sl et al.} (1995) \\
GJ 1194B:  parallax from van Altena {\sl et al.} (1995) \\
LP 229-17:  parallax from (M$_V$, TiO5) relation , spectral type = M3.5 and M$_V$=12.1 (PMSU1)\\
+46 2654: HIP 94701; M$_K$ is consistent with spectral type listed in SIMBAD. \\
+48 3952B: HD 218790B or HIP 114420B. V$\sim 10.4$; 2MASS photometry possibly affected by primary, V=7.4,  
$\Delta \sim 4$\arcsec, $\theta = 157^o$ \\
G 273-93: parallax from (M$_V$, TiO5) relation,  spectral type = M2 and M$_V$=10.3 (PMSU1)\\
\end{table}

\clearpage
\begin{table}
\begin{center}
{\bf Table 2 } \\
{Close binary stars in the final sample}
\begin{tabular}{rrrrccrccrccr}
\tableline\tableline
NLTT & Name & $\alpha$ (2000) & $\delta$ & $m_r$ &  J & H & K$_S$ & $\pi$ & ref.\\
\tableline
 +45:4408A & Gl  4A &  00 \ 05 \ 40.8& 45 \ 48 \ 37&   8.9&  6.161&\nodata&  5.291&   88.6 $\pm$   2.3 &    2 \\
+45:4408B & Gl  4B &  00 \ 05 \ 40.9& 45 \ 48 \ 43&   9.1&  6.117&\nodata&  5.259&  88.6 $\pm$   2.3 &    2 \\
 764- 87  & GJ 1005AB& 00 \ 15 \ 27.9&-16 \  08 \ 00 &  10.2& 7.205&  6.702&  6.394&  182.1 $\pm$ 6.8 &  2 \\ 
 -21:1051 & Gl 185AB  &    05 \ 02 \ 28.4&-21 \ 15 \ 23&   8.1&\nodata&\nodata&  4.586&  117.4 $\pm$   1.8 &  1 \\ 
 +32:1582 & Gl 278C$^1$    &  07 \ 34 \ 37.4& 31 \ 52 \ 10&   9.1& 6.086&\nodata&  5.224&   74.7 $\pm$   2.5 &    2 \\ 
 +15:1957B & GJ 1120B  &  09 \ 01 \ 17.5& 15 \ 15 \ 57&   9.5&  6.687&  6.027&  5.914&   54.6 $\pm$   3.2 &    1 \\ 
R948/LP735-11 & &   12 \ 28 \ 53.0&-10 \ 39 \ 50&  11.0& 7.619&  7.073&  6.788&  \nodata &      \\ 
 G 63-36/LP438-8 & Gl 516 AB &   13 \ 32 \ 44.5& 16 \ 48 \ 39&  11.0& 7.643&  7.048&  6.817&   61.5 $\pm$  5.7 &    2 \\
 +47:2112B & Gl 537B & 14 \ 02 \ 33.1& 46 \ 20 \ 23&  10.1&  6.205&  5.636&  5.397&   88.8 $\pm$  3.9 &  2.5 \\ 
 +47:2112A & Gl 537A& 14 \ 02 \ 33.2& 46 \ 20 \ 26&  10.0&  6.325&  5.637&  5.401&   88.8 $\pm$   3.9 &    2 \\
 W 1225   & Gl 856AB &     22 \ 23 \ 29.0& 32 \ 27 \ 33&  11.4&  6.905&  6.300&  6.048&   62.2 $\pm$  10.0    1 \\ 
  Gl 888AB & G 216-26   & & 23 \ 06 \ 02.2& 42 \ 19 \ 45&  11.3&  8.560&  7.900&  7.751&   17.5 $\pm$   7.6 &    1 \\
 -17:6768 & Gl 897AB & &    23 \ 32 \ 46.5&-16 \ 45 \  08&  10.8&  6.691&  6.070&  5.832&  89.9 $\pm$   7.3 &    1 \\ 
\tableline\tableline
\end{tabular}
\end{center}
\tablecomments{
1. Gl 278C is the well-known eclipsing binary, YY Geminorum.}
\end{table}

\begin{deluxetable}{rrrrrrrrrrcrrrrrc}
\rotate
\tabletypesize{\scriptsize}
\tablecolumns{16}
\tablenum{3}
\tablewidth{0pt}
\tablecaption{Photometry and astrometry of NLTT stars}
\tablehead{
\colhead{NLTT} & \colhead {Name} & \colhead {LHS} &
\colhead{$\alpha$ (2000)} & \colhead {$\delta$} & \colhead {m$_r$} &
\colhead{V } & \colhead {B-V } & \colhead {V-R } &
\colhead{V-I } & \colhead {Ref$_{ph}$ } & \colhead {J } &
\colhead{H } & \colhead {K$_S$ } & \colhead {$\pi_{trig}$ } & \colhead {Ref$_\pi$ } }
\startdata
 +44:4548 & Gl   2  &1014&  00 \ 05 \ 10.7& 45 \ 47 \ 11&   9.8&   9.95&   1.51&   0.97&   2.08&   1&  6.695&  6.093&  5.848&   87.0 $\pm$   1.4 &    1 \\ 
   G 158-27 & GJ 1002 & 2&  00 \ 06 \ 43.2& -7 \ 32 \ 14&  13.0&  13.83& \nodata&   1.59&   3.59&  11&  8.351&  7.786&  7.448&  213.0$\pm$ 3.6 &    2 \\ 
 -27:16  & Gl   7  &1026&   00 \ 09 \ 04.2&-27 \  07 \ 19&  11.5&  11.68&   1.47&   0.95&   1.90&  11&  8.642&  8.142&  7.851&   42.8 $\pm$   2.6 &    1 \\ 
 464- 42  & Gl  12  &1050&  00 \ 15 \ 49.1& 13 \ 33 \ 21&  12.2&  12.60&   1.65&   1.15&   2.56&  11&  8.615&  8.059&  7.789&   86.6 $\pm$   13.4 &    2 \\ 
 404- 61  & GJ 1006A & 107& 00 \ 16 \ 14.5& 19 \ 51 \ 38&  10.9&  12.26&   1.54&   1.21&   2.79&   1&  7.900&  7.311&  7.100&   66.1 $\pm$  1.6 &    2 \\ 
 404- 62 & GJ 1006B & 108&  00 \ 16 \ 16.0& 19 \ 51 \ 51&  12.1&  13.21&   1.58&   1.21&   2.81&   1&  8.907&  8.321&  8.104&   66.1 $\pm$   1.6 &    2 \\ 
  G 158-52 & LTT141 & &     00 \ 17 \ 40.8& -8 \ 40 \ 55&  11.2&  11.02&   1.43&  0.89& 1.84&   1&  8.082&  7.467&  7.248&   28.4 $\pm$   2.3 &    1 \\ 
 +43: 44B & Gl  15B &   4&  00 \ 18 \ 25.5& 44 \  01 \ 37&  10.3&  11.06&   1.82&   1.24&   2.82&   1&  6.793&  6.184&  5.952&  282.0 $\pm$  2.2 &    2 \\
\enddata
\tablecomments{ 
Column 1 lists the designation from the NLTT catalogue: R=Ross, W=Wolf, Oxf=Oxford catalogue. We
have added Lowell Observatory Proper Motion Survey identifications (Giclas {\sl et al.}, 1971) where appropriate. \\
Column 2 lists an alternative name, usually from the pCNS3; 
Column 3 gives the LHS number; \\
Columns 4 and 5 list the position of the 2MASS source; \\
Column 6 lists the r-band photometry from the NLTT catalogue; \\
Columns 7-10 list the optical photometry, and column 11 gives the source:
1 - Weis (1984, 1986, 1987, 1988, 1991, 1993, 1996, 1999); 
3 - Hipparcos (ESA, 1997); 4 - pCNS3; 5 - Sandage \& Kowal (1986); 6 - Ryan (1989, 1992); 7 - Fleming (1998); 
8 - Patterson {\sl et al.} (1998); 9 - Eggen (1966, 1975, 1980, 1987); 10 -Andruk {\sl et al.} (1995); 11 - Bessell (1990); 
12 - Leggett (1992); 13 - Dawson \& Forbes (1989, 1992); 14 - Humphreys {\sl et al.} (1991); 15 - Gullixson  {\sl et  al.}
(1995); 16 - USNO, Harrington {\sl et al.} (1993) and refs within (BV only); 17 - Hartwick {\sl et al.} (1984); 
18 - Dahn {\sl et al.}, 2000; 19 - Stauffer \& Hartmann (1986); 20 - Upgren \& Lu (1986); 21 - Reid (1990); 
22 - Kron {\sl et al.}, 1957; 23 - van Altena {\sl et al.} (1995); 24 - Laing (1989).  \\
Columns 12-14 list the 2MASS photometry; \\
Column 15 lists the trigonometric parallax, if available, and column 16
gives the source of the astrometry: 1 - Hipparcos (ESA, 1997); 1.5 - Hipparcos data for
the primary in a multiple system; 2 - Yale catalogue, van Altena {\sl et al.}
(1995); 2.5 - Yale catalogue data for the primary star in a multiple system; 3 - Tinney (1996); 
4 - Dahn {\sl et al.}, 2000\\}
\end{deluxetable}

\begin{deluxetable}{rcccccrrrr}
\rotate
\tabletypesize{\scriptsize}
\tablecolumns{10}
\tablenum{4}
\tablewidth{0pt}
\tablecaption{Distance Modulus Estimates for NLTT stars}
\tablehead{
\colhead{NLTT} & \colhead {(m-M)$_{V-K}$} & \colhead {(m-M)$_{V-I}$} &
 \colhead {(m-M)$_{I-J}$} & \colhead {(m-M)$_{ph}$} &\colhead {(m-M)$_\pi$} & 
\colhead {d$_f$(pc)} & \colhead{M$_K$} & \colhead{M$_V$} & \colhead{ 20pc?} } 
\startdata
 +44:4548*   &   0.24$\pm$0.41&   0.20$\pm$0.40&   0.26$\pm$0.42&   0.23$\pm$0.24&   0.30$\pm$0.04&   11.5$\pm$0.2&   5.55&   9.65& Y   \\ 
 +45:4408A   &   0.17$\pm$0.41&   0.27$\pm$0.40&   0.25$\pm$0.42&   0.23$\pm$0.24&   0.26$\pm$0.06&   11.3$\pm$0.3&   5.03&   8.71& Y   \\ 
 +45:4408B*  &   0.05$\pm$0.41&   0.33$\pm$0.40&  -0.15$\pm$0.42&   0.08$\pm$0.24&   0.26$\pm$0.06&   11.1$\pm$0.3&   5.03&   8.79& Y   \\ 
   G 158-27  &  -1.36$\pm$0.41&  -1.49$\pm$0.22&  -1.35$\pm$0.31&  -1.42$\pm$0.16&  -1.64$\pm$0.04&    4.7$\pm$0.1&   9.07&  15.45& Y*  \\ 
 -27:16*     &   2.57$\pm$0.41&   2.53$\pm$0.40&   2.37$\pm$0.42&   2.49$\pm$0.24&   1.84$\pm$0.14&   25.6$\pm$1.4&   5.81&   9.64& N   \\ 
 764- 87     &  -0.74$\pm$0.41&  -0.94$\pm$0.75&  -1.52$\pm$0.70&  -1.01$\pm$0.32&  -1.30$\pm$0.08&    5.6$\pm$0.2&   7.63&  12.74& Y*  \\ 
 464- 42     &   1.13$\pm$0.41&   1.25$\pm$0.40&   1.11$\pm$0.42&   1.17$\pm$0.24&   0.31$\pm$0.37&   14.3$\pm$1.5&   7.01&  11.82& Y   \\ 
 404- 61     &  -0.11$\pm$0.41&   0.10$\pm$0.40&  -0.74$\pm$0.70&  -0.17$\pm$0.27&   0.90$\pm$0.05&   14.1$\pm$0.3&   6.36&  11.52& Y   \\
\enddata
\tablecomments{ 
Column 1 lists the designation from the NLTT catalogue, adding Lowell Observatory identifications; \\
Column 2 gives the distance modulus derived from the (V-K$_S$) photometric parallax; Columns 3 and 4 list
distance moduli for stars with I-band photometry, based on (V-I) and (I-J) respectively; Column 5
gives the weighted average of the photometric parallax measurements; Column 6 lists the
distance modulus indicated by the trigonometric parallax; \\
Column 7 gives our final estimate of the distance, based on a weighted average of the photometric
average and the trigonometric result; \\ Columns 8 and 9 list the resultant absolute magnitudes
at K and V, respectively; \\ Column 10 indicates whether the star lies within our distance limit of
20 parsecs (Y), within 1$\sigma$ of the boundary (?) or beyond the limit (N). A
(+) indicates that the star is not included in the pCNS3. }
\end{deluxetable}

\clearpage

\begin{table}
\begin{center}
{\bf Table 5 } \\
{Comparison between photometric and trigonometric distance moduli}
\begin{tabular}{rccrccrccrccr}
\tableline\tableline
$\langle M_V \rangle$ & $\Delta_{\pi 1}$ & $\sigma$ & n$_1$ &$\Delta_{\pi 2}$ & $ \sigma$ & n$_2$ &
$\Delta_{\pi 3}$ & $ \sigma$ & n$_3$ &$\Delta_{\pi 4}$ & $ \sigma$ & n$_4$ \\
\tableline
     7.0&  0.239&  0.289&   19&  0.065&  0.279&    9&  0.286&  0.465&    9&  0.225&  0.302&   19 \\
     9.0&  0.178&  0.431&   87&  0.179&  0.423&   70&  0.172&  0.499&   70&  0.174&  0.415&   87 \\
    11.0&  0.075&  0.542&   83&  0.075&  0.417&   75&  0.089&  0.758&   74&  0.082&  0.518&   83 \\
    13.0& -0.211&  0.608&   36&  0.032&  0.667&   34&  0.318&  0.524&   33& -0.013&  0.636&   36 \\
    15.0&  0.046&  0.413&   15&  0.135&  0.335&   13&  0.159&  0.407&   13&  0.092&  0.350&   15 \\
    17.0& -0.128&  0.150&    6&  0.016&  0.082&    5& -0.090&  0.220&    5& -0.083&  0.120&    6 \\
\tableline\tableline
\end{tabular}
\end{center}
Mean residuals, as a function of absolute magnitude, between photometric and trigonometric parallax 
estimates for stars with trigonometric parallaxes accurate to better than 10\%. The
residuals are listed as differences in distance modulus in the sense
\begin{displaymath}
\Delta \ = \ \Sigma ( (m-M)_\pi - (m-M)_{phot} ) / n
\end{displaymath}
where (m-M)$_{phot}$ is derived from the photometric parallax; $\sigma$ is the dispersion about the
mean; n is number of stars contributing to each bin.  The table lists comparisons against
the individual photometric parallaxes ($\Delta_{\pi 1}$ against (m-M)$_{V-K}$,  $\Delta_{\pi 2}$ against (m-M)$_{V-I}$
and $\Delta_{\pi 3}$ against (m-M)$_{I-J}$), and against the weighted average of the photometric
estimates ($\Delta_{\pi 4}$).
\end{table}

\clearpage
\begin{table}
\begin{center}
{\bf Table 6 } \\
{Spectroscopically-confirmed ultracool dwarfs}
\begin{tabular}{rrcrcrrrccr}
\tableline\tableline
NLTT &  $\alpha$ (2000) & $\delta$ & m$_r$ & I/Sp. & J & H & K$_S$ & d (pc) & ref & M$_K$ \\
\tableline
368-128     &     09 \   00 \  23.5& 21 \  50 \ 05&  15.5&M6.5&  9.423&  8.856&  8.429&$5.2\pm1$&   1& 9.85  \\ 
 860- 46     &    15 \  53 \  57.1&-23 \  11 \  52&  16.3&13.56& 11.570& 10.957& 10.636&$22.2\pm4$&   2 & 8.90\\ 
 213- 67     &   10 \  47 \  12.6& 40 \  26 \  43&  16.3&& 11.417& 10.777& 10.400&$12.7\pm2.5$&   3& 9.88 \\ 
 349- 25   &    00 \ 27 \ 55.9& 22 \  19 \  32&  17.0&M8& 10.608&  9.970&  9.561& $8.4\pm1.7$ & 4 & 9.93\\ 
 315- 53     &    10 \  16 \  34.7& 27 \  51 \  49&  17.4&M7.5& 11.951& 11.294& 10.946&$16.5\pm3$&   4 & 9.86 \\ 
 944- 20               &  03 \  39 \  35.2&-35 \  25 \  44&  17.5&M9.5& 10.748& 10.017&  9.525&  $5.0\pm0.1$&   5 & 8.02 \\ 
 356-770     &      03 \  30 \   05.0& 24 \ 05 \  28&  18.1&M7& 12.357& 11.745& 11.361& $20.2\pm4$ &  4 &9.83 \\ 
 213- 68    &    10 \  47 \  13.8& 40 \  26 \  49&  18.7&& 12.445& 11.705& 11.277& $16.3\pm3$ & 3 & 10.22\\ 
 413- 53     &     03 \ 50 \ 57.3& 18 \  18 \   6&  19.2&M9& 12.951& 12.222& 11.763& 19.9 $\pm$ 4.0 & 4 & 10.27 \\ 
\tableline\tableline
\end{tabular}
\end{center}
Column 5 lists either Cousins I-band photometry or the spectral type. \\
References: \\
1. Scholz {\sl et al.} (2001), LHS 2090 - distance from (J-K$_S$) \\
2. Ardila {\sl et al.} (2001), UScoCTIO 4 - distance from (I-J) \\
3. Gizis {\sl et al.} (1999) - distance from (J-K$_S$) \\
4. Gizis {\sl et al.} (2000) - distance from (J-K$_S$); LP 315-53 = LHS 2243 \\
5. Tinney (1996, 1998) - distance from trigonometric parallax 
\end{table}
\clearpage
\begin{table}
\begin{center}
{\bf Table 7 } \\
{Photometrically-selected ultracool dwarfs}
\begin{tabular}{rrrcrrrrrcr}
\tableline\tableline
NLTT & LHS & $\alpha$ (2000) & $\delta$ & $m_r$ &  J & H & K$_S$ & M$_K$ & d$_{J-K}$ (pc) & d$_f$ \\
\tableline
   G118-43  & & 10 \ 15 \ 06.9& 31 \ 25 \ 11&  12.9&  9.410&  8.780&  8.410&   9.84&   5.2 $\pm$ 1.0 &17.7 \\
   G180-11  & & 15 \  55 \  31.8& 35 \  12 \   02&  12.9&  8.999&  8.290&  7.986&  9.87&   4.2 $\pm$   0.8 & 13.3\\ 
  G139-3    & & 16 \  58 \  25.3& 13 \  58 \  10&  13.5&  8.859&  8.284&  7.737& 10.12&   3.3 $\pm$   0.7 & 13.6\\ 
 G199-16    &  6234 & 12 \  29 \   09.5& 62 \  39 \  38&  14.2& 10.337&  9.775&  9.315&  9.89&   7.7 $\pm$   1.5 \\ 
 245- 10     &  1378    & 02 \  17 \   09.9& 35 \  26 \  33&  14.7&  9.965&  9.355&  8.974&   9.82&   6.8 $\pm$   1.4& 10.2 \\ 
 645- 53     &               & 00 \  35 \  44.1& -05 \  41 \  10&  14.9& 10.717& 10.084&  9.716&  9.85&   9.4 $\pm$   1.9 \\ 
 714- 37     &               & 04 \  10 \  48.0&-12 \  51 \  42&  15.1& 11.060& 10.469& 10.015&  9.94&  10.3 $\pm$   2.1 \\ 
 649- 72     &   1363& 02 \  14 \  12.5& -03 \  57 \  43&  15.5& 10.472&  9.839&  9.466&  9.86&   8.4 $\pm$   1.7 \\ 
 264- 45     &            & 11 \  22 \  42.7& 37 \  55 \  48&  16.0& 11.302& 10.656& 10.305&   9.84&  12.4 $\pm$   2.5 \\
 740- 20     &            & 14 \  31 \  15.6&-13 \  18 \  24&  16.1& 11.136& 10.496& 10.121&  9.88&  11.2 $\pm$   2.2 \\ 
 423- 31     &            & 07 \  52 \  23.9& 16 \  12 \  15&  16.3& 10.831& 10.192&  9.819&  9.87&   9.8 $\pm$   2.0 \\ 
 655- 48     &            & 04 \  40 \  23.2& -05 \  30 \ 08&  16.4& 10.681&  9.985&  9.557& 10.12&   7.7 $\pm$   1.5 \\ 
 914- 54     &  3003    & 14 \  56 \  38.3&-28 \   09 \  47&  16.4&  9.957&  9.327&  8.917&  9.93&   6.3 $\pm$   1.3 & 6.6\\ 
 651- 17     &  1450& 02 \  50 \   02.3& -08 \   08 \  41&  16.5& 11.878& 11.226& 10.850&  9.91&  15.4 $\pm$   3.1 \\ 
 593- 68     &  1604& 03 \  51 \   00.0&  00 \  52 \  44&  16.5& 11.262& 10.592& 10.191& 10.00&  10.9 $\pm$   2.2 & 14.4\\ 
 800- 58     &            & 14 \  25 \  13.3&-16 \  24 \  56&  16.6& 11.469& 10.918& 10.478&   9.82&  13.5 $\pm$   2.7 & \\ 
 698- 2      &            & 21 \  32 \  29.7& -05 \  11 \  58&  16.6& 11.439& 10.715& 10.385&  9.96&  12.1 $\pm$   2.4 \\ 
 763-  3     &            & 23 \  37 \  38.3&-12 \  50 \  27&  16.7& 11.461& 10.851& 10.427&  9.92&  12.6 $\pm$   2.5 \\ 
 927- 32     &  3566& 20 \  39 \  23.8&-29 \  26 \  33&  16.7& 11.346& 10.768& 10.352&   9.83&  12.7 $\pm$   2.5 \\ 
 785-  4     &            & 08 \  24 \  29.3&-19 \  37 \  36&  16.8& 11.896& 11.312& 10.905&   9.82&  16.5 $\pm$   3.3 \\ 
 985- 98     &            & 23 \   09 \  14.2&-35 \  31 \  59&  16.9& 12.035& 11.351& 10.986&  9.95&  16.1 $\pm$   3.2 \\ 
 335- 12     &            & 18 \  39 \  33.0& 29 \  52 \  16&  17.2& 10.964& 10.381&  9.960&  9.85&  10.5 $\pm$   2.1 \\ 
 775- 31     &         & 04 \  35 \  16.1&-16 \   06 \  57&  17.4& 10.396&  9.780&  9.336&  9.98&   7.4 $\pm$   1.5 \\ 
 218-  8     &  2645& 12 \  53 \  12.4& 40 \  34 \   03&  17.5& 12.177& 11.557& 11.173&  9.85&  18.4 $\pm$   3.7 \\ 
 441- 34    & 3002  & 14 \  56 \  27.8& 17 \  55 \   07&  17.5& 11.931& 11.320& 10.936&   9.83&  16.6 $\pm$   3.3 & 19.4 \\ 
 718-  5     &           & 05 \  35 \  21.2& -09 \  31 \  06&  17.5& 11.851& 11.201& 10.814&  9.93&  15.1 $\pm$   3.0 \\ 
 220- 13     &            & 13 \  56 \  41.4& 43 \  42 \  58&  17.5& 11.704& 11.031& 10.634& 10.00&  13.4 $\pm$   2.7 \\ 
 229- 30     & 3406   & 18 \  43 \  22.1& 40 \  40 \  21&  17.5& 11.299& 10.667& 10.269&  9.91&  11.8 $\pm$   2.4 & 14.1 \\ 
 789- 23     &            & 10 \   06 \  31.9&-16 \  53 \  26&  17.6& 12.041& 11.421& 11.000&  9.94&  16.3 $\pm$   3.3 \\ 
 647- 13     &           & 01 \   09 \  51.1& -03 \  43 \  26&  17.9& 11.695& 10.921& 10.418& 10.47&   9.8 $\pm$   2.0 \\ 
 666-  9     &    2065& 08 \  53 \  36.2& -03 \  29 \  32&  17.9& 11.185& 10.468&  9.972& 10.32&   8.5 $\pm$   1.7 & 12.8 \\ 
 763- 38     &            & 23 \  37 \  14.9& -08 \  38 \  08&  18.0& 12.246& 11.603& 11.206&  9.93&  18.0 $\pm$   3.6 \\ 
 267-299    &            & 12 \  52 \  17.0& 33 \  57 \  39&  18.1& 12.246& 11.601& 11.239&  9.86&  18.9 $\pm$   3.8 \\ 
 429- 12     &   2215& 09 \  59 \  56.0& 20 \  02 \  34&  18.1& 12.244& 11.615& 11.196&  9.95&  17.7 $\pm$   3.5 \\ 
 423- 14     &   1937& 07 \  41 \  06.8& 17 \  38 \  45&  18.1& 11.995& 11.362& 10.969&  9.90&  16.3 $\pm$   3.3 \\ 
\tableline\tableline
\end{tabular}
\end{center}
\end{table}
\clearpage
\begin{table}
\begin{center}
{\bf Table 6 (contd.) } \\
{Photometrically-selected ultracool dwarfs}
\begin{tabular}{rrrcrrrrrcr}
\tableline\tableline
NLTT & LHS & $\alpha$ (2000) & $\delta$ & $m_r$ &  J & H & K$_S$ & M$_K$ & d$_{J-K}$ (pc) & d$_f$ \\
\tableline
 658-106     &            & 05 \  37 \  23.3& -08 \  16 \  05&  18.2& 12.305& 11.675& 11.304&  9.85&  19.6 $\pm$   3.9 \\ 
 888- 18     &            & 03 \  31 \  30.2&-30 \  42 \  38&  18.2& 11.371& 10.699& 10.276& 10.06&  11.1 $\pm$   2.2 \\ 
 890-  2     &            & 04 \  13 \  39.8&-27 \  04 \  29&  18.4& 12.214& 11.578& 11.190&  9.90&  18.1 $\pm$   3.6 \\ 
 859-  1     &            & 15 \  04 \  16.2&-23 \  55 \  56&  18.4& 12.025& 11.389& 11.031&   9.83&  17.4 $\pm$   3.5 \\ 
 754- 14     &            & 20 \  04 \  18.4&-12 \  20 \  31&  18.9& 12.827& 12.153& 11.829&   9.84&  25.0 $\pm$   5.0 \\ 
 695-351     &            & 20 \  41 \  41.0& -03 \  33 \  53&  19.0& 12.528& 11.882& 11.504&  9.90&  21.0 $\pm$   4.2 \\ 
 649- 93     &            & 02 \  18 \  57.8& -06 \  17 \  49&  19.2& 12.920& 12.186& 11.860&  9.98&  23.8 $\pm$   4.8 \\ 
\tableline\tableline
\end{tabular}
\end{center}
Column 10 lists the distance based on the (M$_K$, (J-K$_S$)) calibration given in the text; Column
11 lists d$_f$ from Table 2 for stars with optical photometry.
\end{table}

\clearpage
\begin{figure}
\caption{ Aitoff projections, centred on ($\alpha=12$ hours, $\delta=0^o$), showing the distribution 
of NLTT stars on the celestial sphere; grid lines are at Right Ascension 4, 8, 12, 16 and 20 hours.
The transition from the Palomar Schmidt survey to the Bruce proper
motion data at $\delta = -33^o$ is clearly evident at both intermediate and faint magnitudes, as 
is the location of the Galactic Plane.  }
\end{figure}

\begin{figure}
\plotone{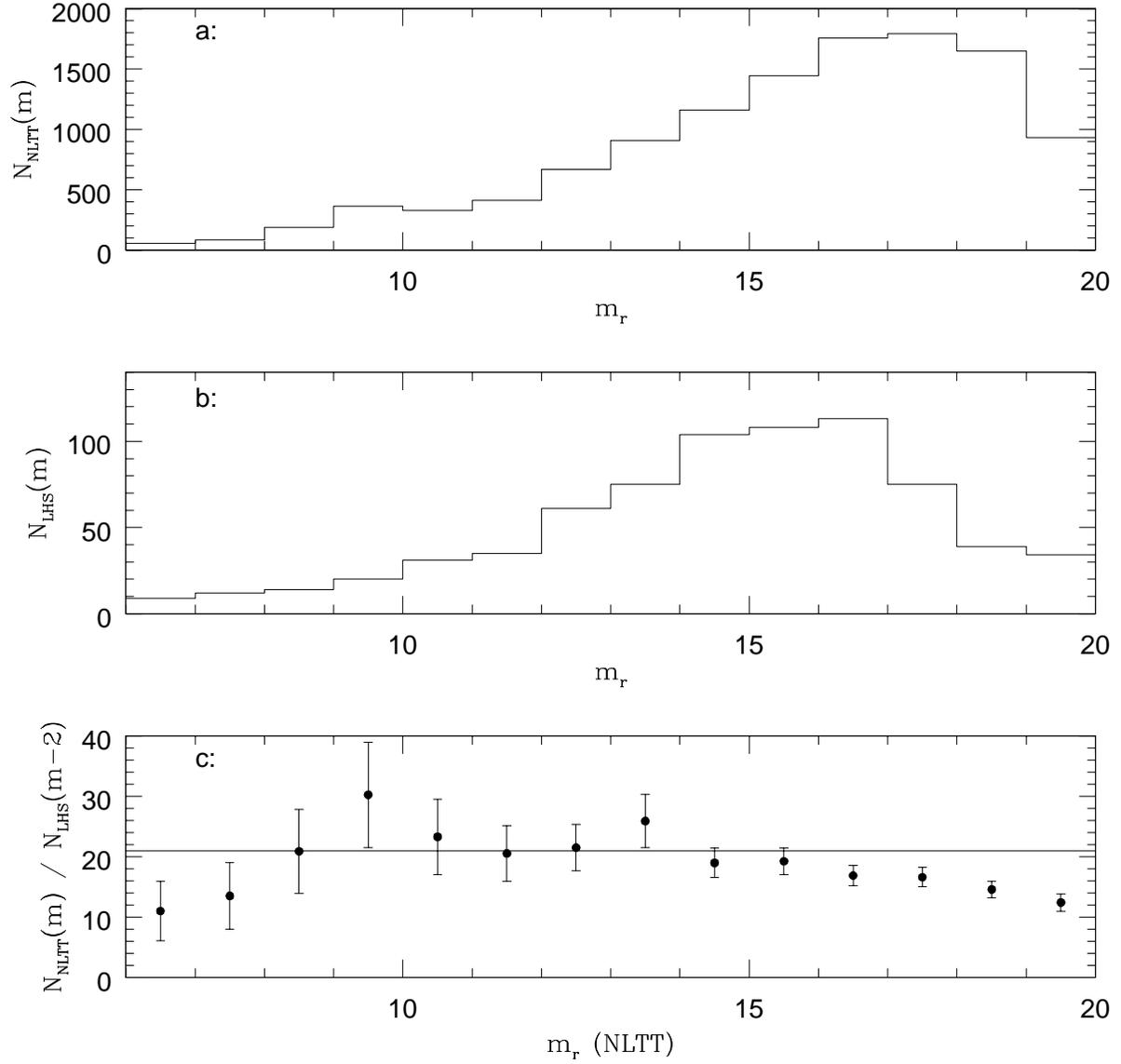}
\caption{ Number counts at high galactic latitude from a: the NLTT survey, and b: the LHS
survey. Both datasets are drawn from (10 $< \alpha < 16$ hours; $-20^o < \delta < +50^o$).
c: the ratio between NLTT and LHS number counts, making allowance for the different
sampling volumes as described in the text.} 
\end{figure}

\begin{figure}
\caption{ Aitoff projections, centred on ($\alpha=12$ hours, $\delta=0^o$), for NLTT stars in the
area covered by the 2MASS second incremental release (47\% of the sky), 
excluding regions within 10$^o$ of the Plane. The left-hand panels plot the distribution of
bright ($m_r \le 14$) and faint NLTT stars with a 2MASS counterpart within 10\arcsec; the righthand
panels plot the distribution of bright and faint NLTT stars which lie within the same region on
the sky, but lack close ($\Delta < 10\arcsec$) 2MASS counterparts. It is clear that the latter
distribution is not random.}
\end{figure}

\begin{figure}
\plotone{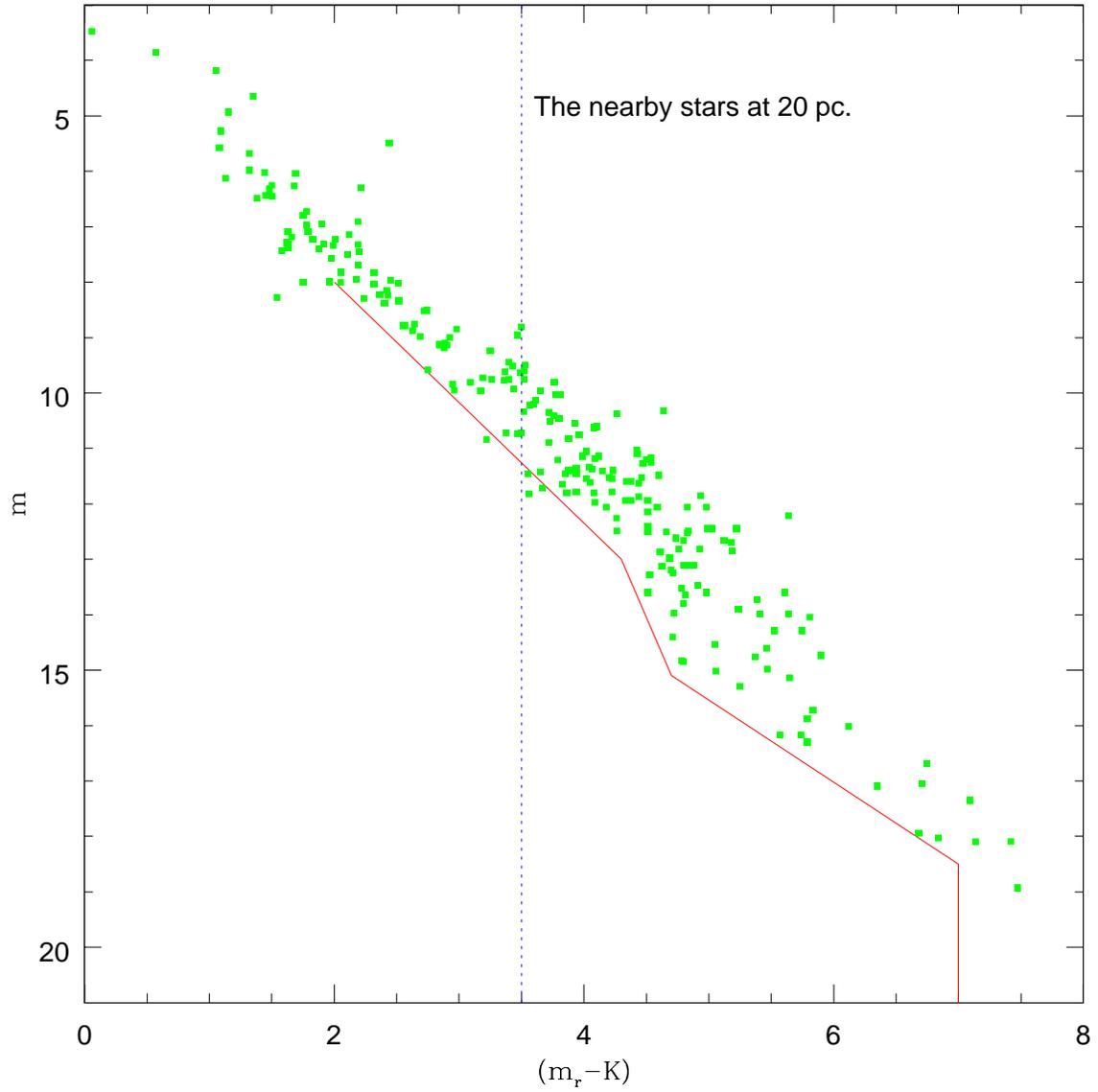}
\caption{ Nearby-star selection in the ($m_r, (m_r, K_s)$) plane. The solid points
plot data for known nearby stars with accurate trigonometric parallax measurements, adjusting
the magnitudes to a distance of 20 parsecs. The solid line underlying the main sequence
outlines the selection criteria described in the test.} 
\end{figure} 

\begin{figure}
\plotone{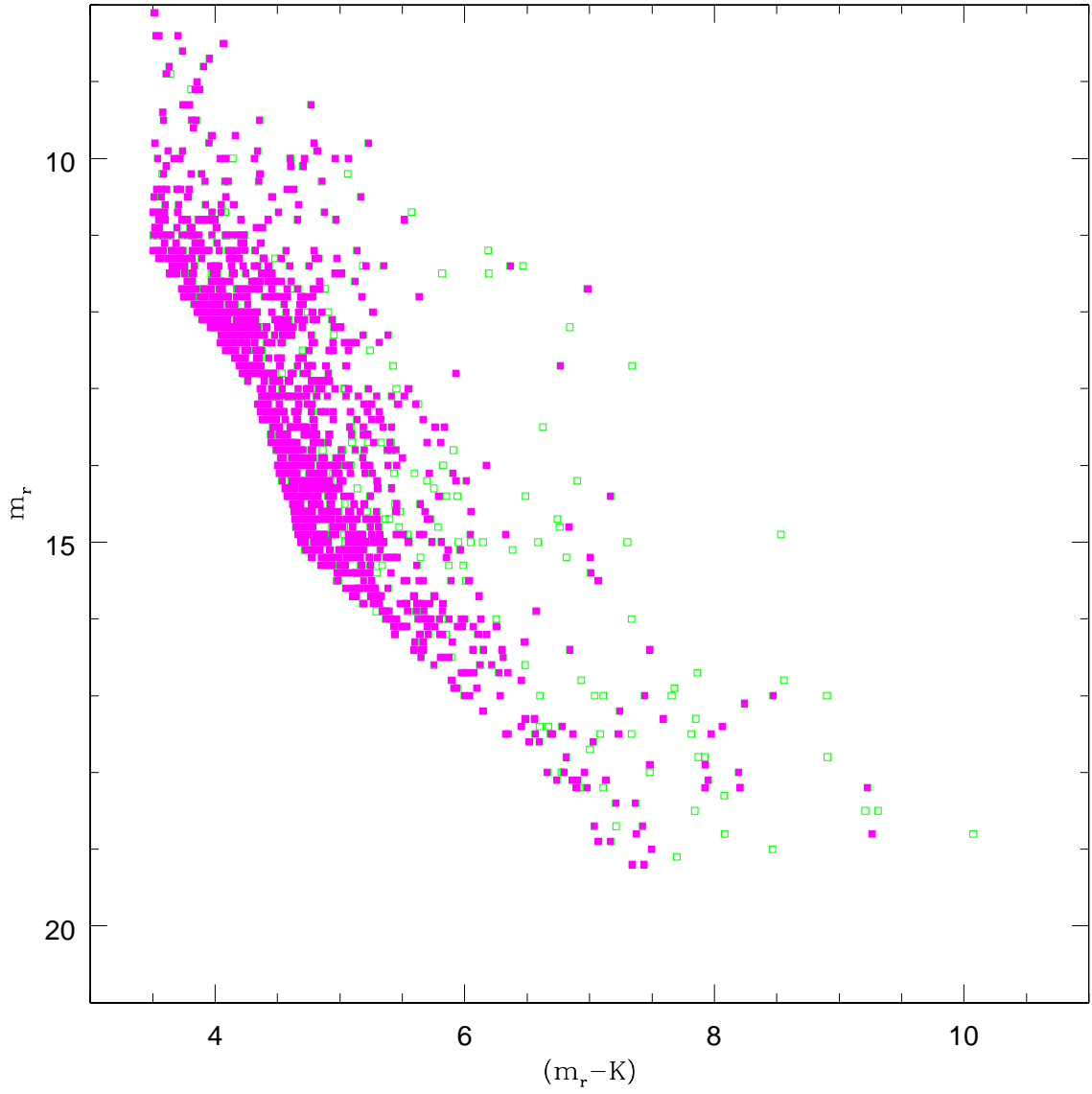}
\caption{The ($m_r, (m_r-K_s)$) colour-magnitude diagram for the NLTT stars in our
primary sample. Open squares identify objects which prove to be mismatches between 
components in cpm binary systems.}
\end{figure}

\begin{figure}
\plotone{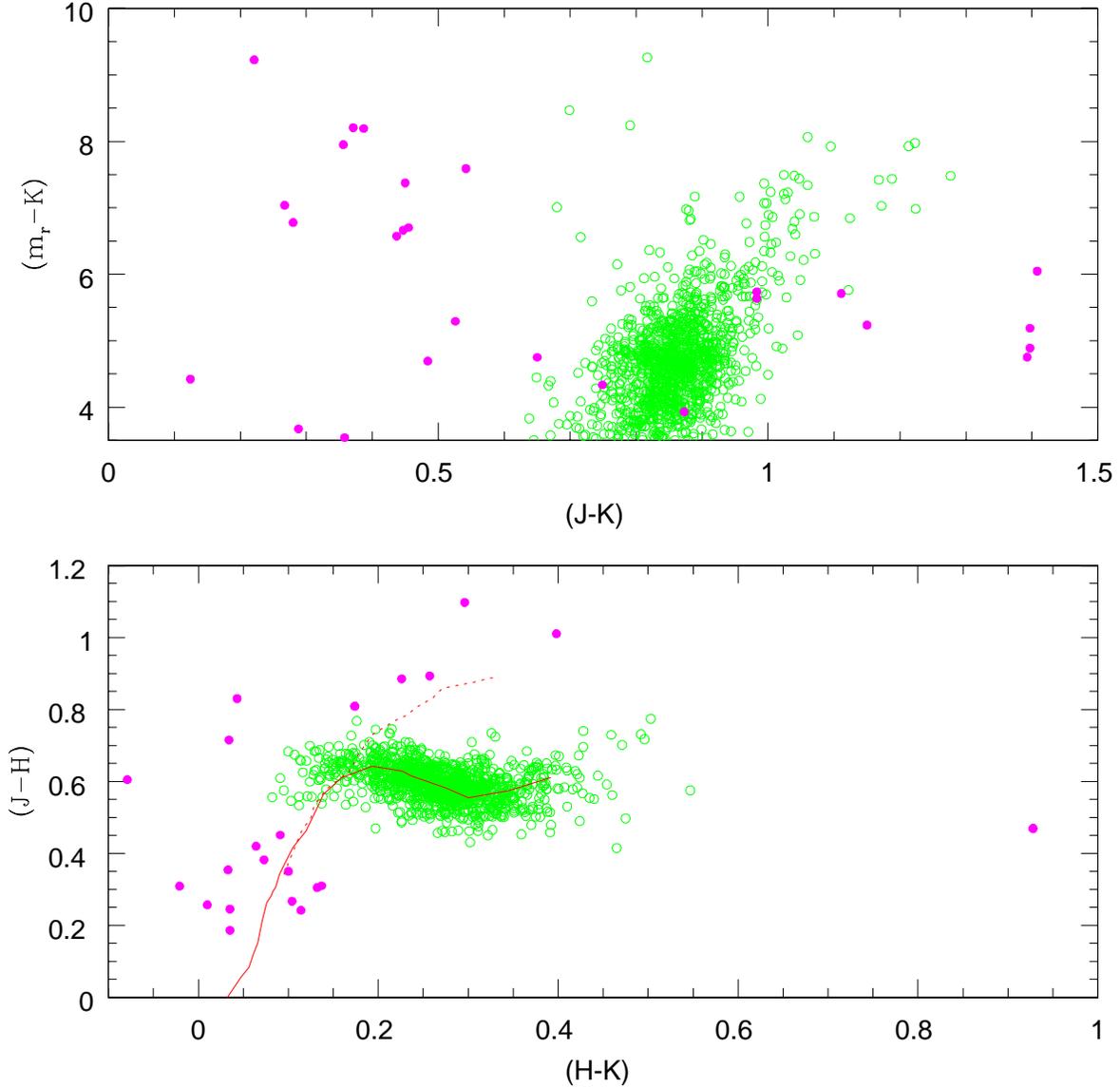}
\caption{The ($(m_r-K_s)$, (J-K$_s)$) and ((J-H), (H-K$_s$)) two-colour diagrams
for the 1275-source NLTT sample. The solid line on the latter diagram marks the 
mean main-sequence relation and the dotted line the giant star distribution, both taken
from Bessell \& Brett (1988), transformed to the 2MASS system using the relations
given by Carpenter (2001). Solid points mark 2MASS sources with (J-H)/(H-K)
colours inconsistent with those of M dwarf stars. Several of the 28 outliers have
colours which lie beyond the limits of these diagrams. }
\end{figure}

\begin{figure}
\plotone{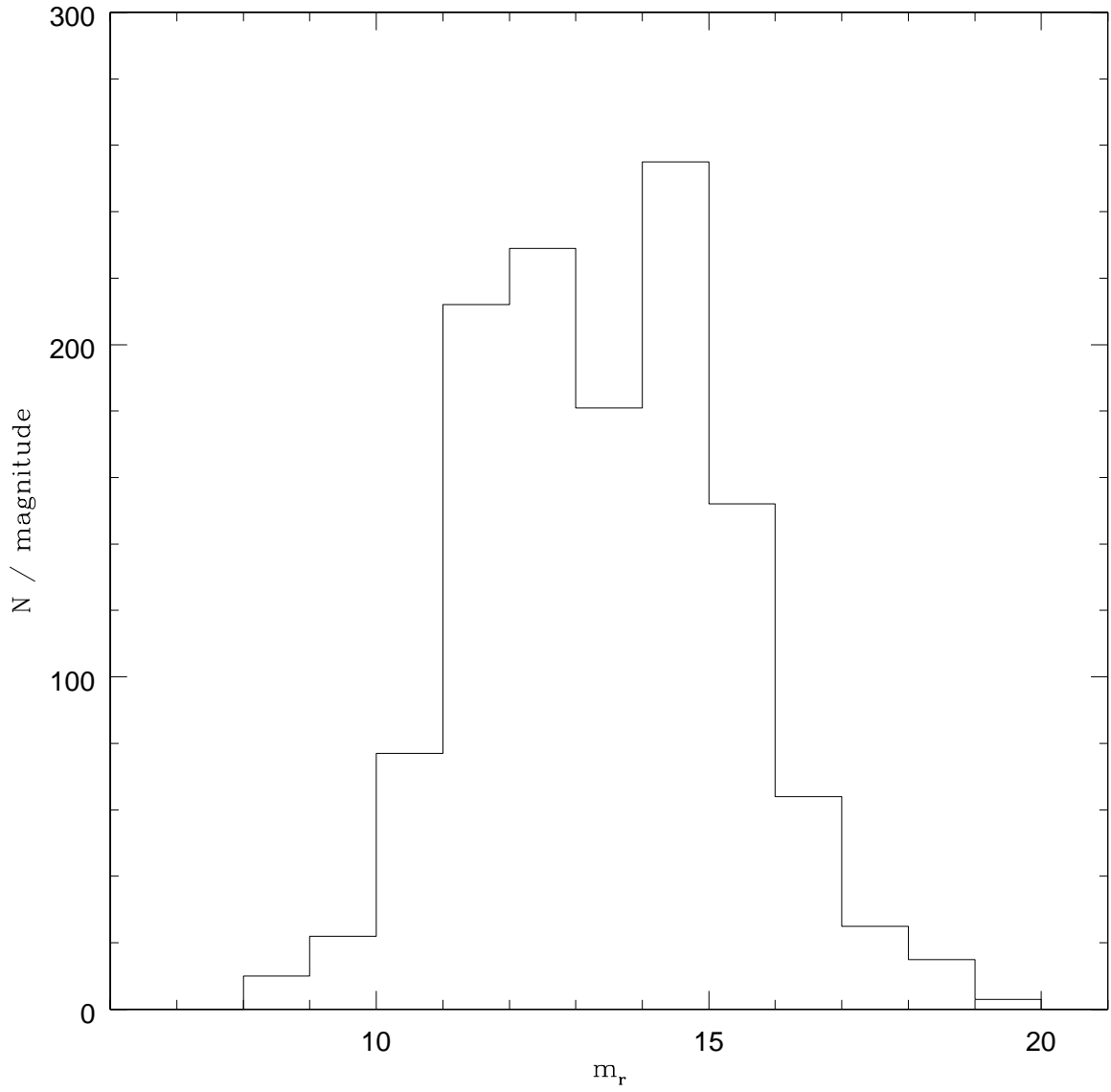}
\caption{The number-magnitude distribution for the 1245 stars in our primary
NLTT sample.}
\end{figure}

\begin{figure}
\plotone{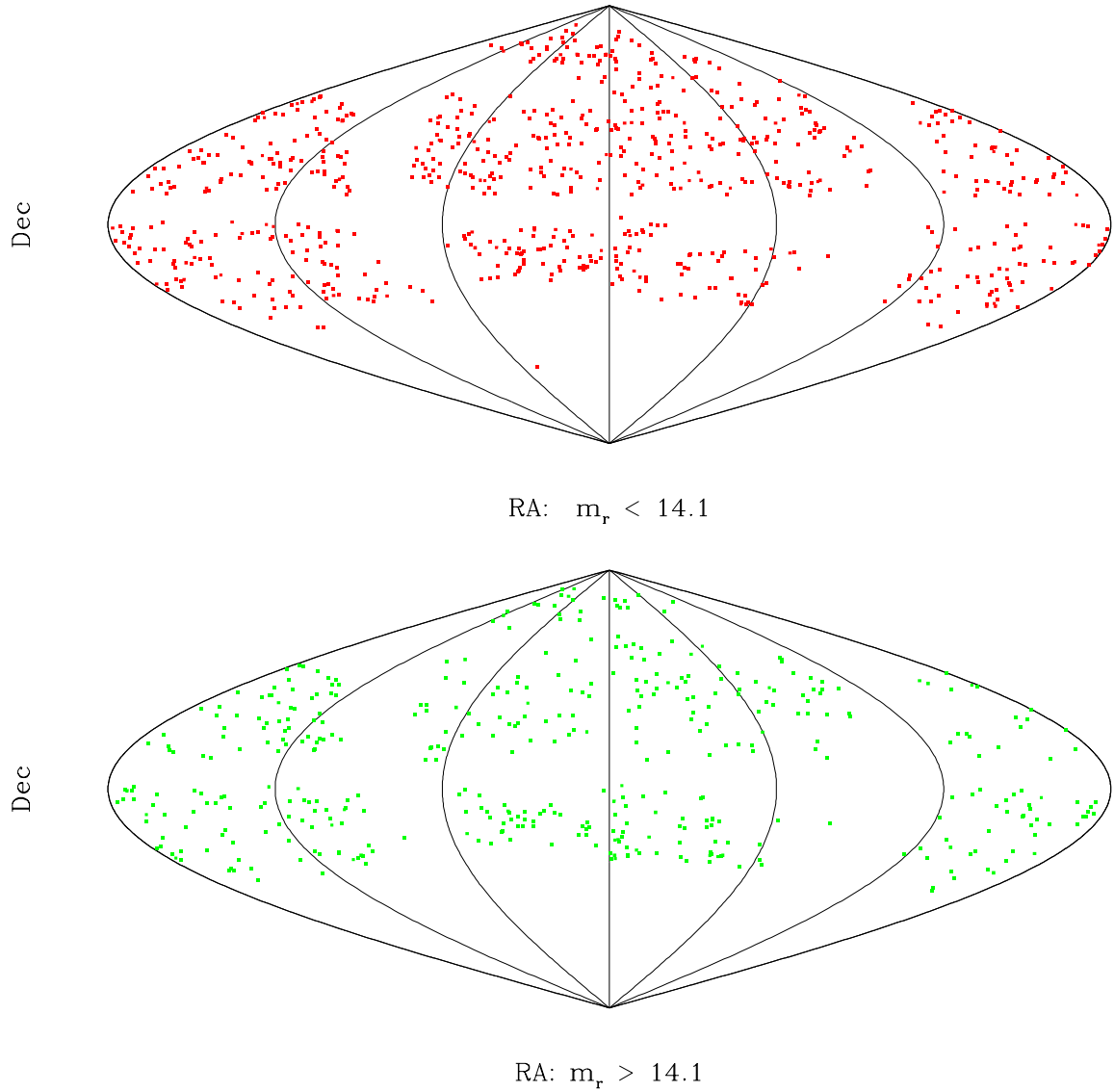}
\caption{Aitoff projections of the ($\alpha, \delta$) distribution of the 1245 stars
in our primary NLTT sample.}
\end{figure}

\begin{figure}
\plotone{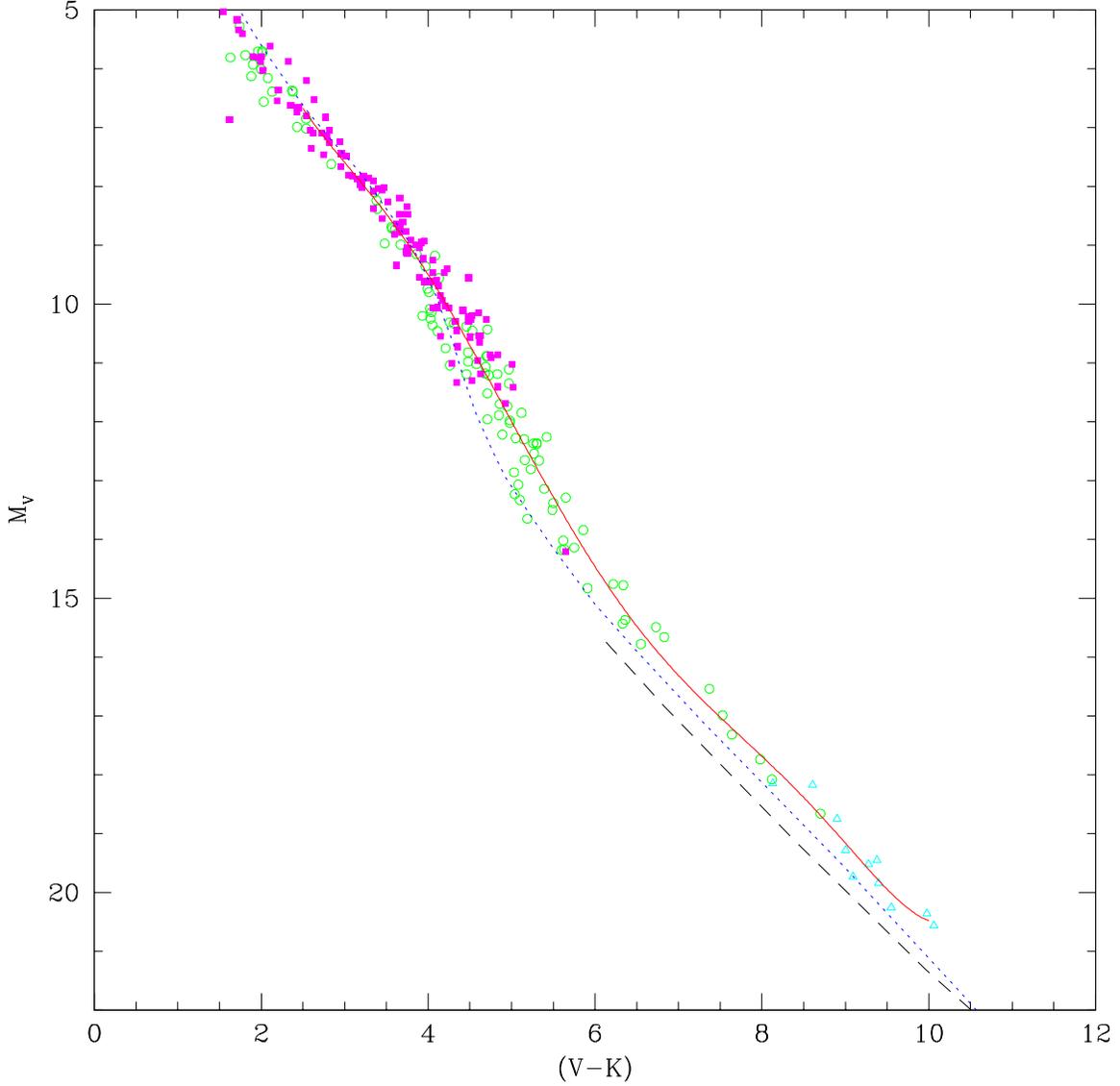}
\caption{The main-sequence in the (M$_V$, (V-K)) plane: open circles are from Leggett's
(1992) compilation, solid squares are CNS2 stars with optical photometry by Bessell (1990)
and 2MASS near-infrared data; open triangles are from Dahn {\sl et al.} (2000). The solid
line marks the best-fit 6th-order polynomial given in the text. Note the steepening of the
main-sequence between M$_V \sim 12$ and $\sim 13.5$. The dotted line plots a 5-Gyr.
isochrone derived from the Baraffe {\sl et al.} (1998) models; the dashed line
plots the 5-Gyr. Dusty model ($M < 0.1M_\odot$) from Chabrier {\sl et al.} (2000).}
\end{figure}

\begin{figure}
\plotone{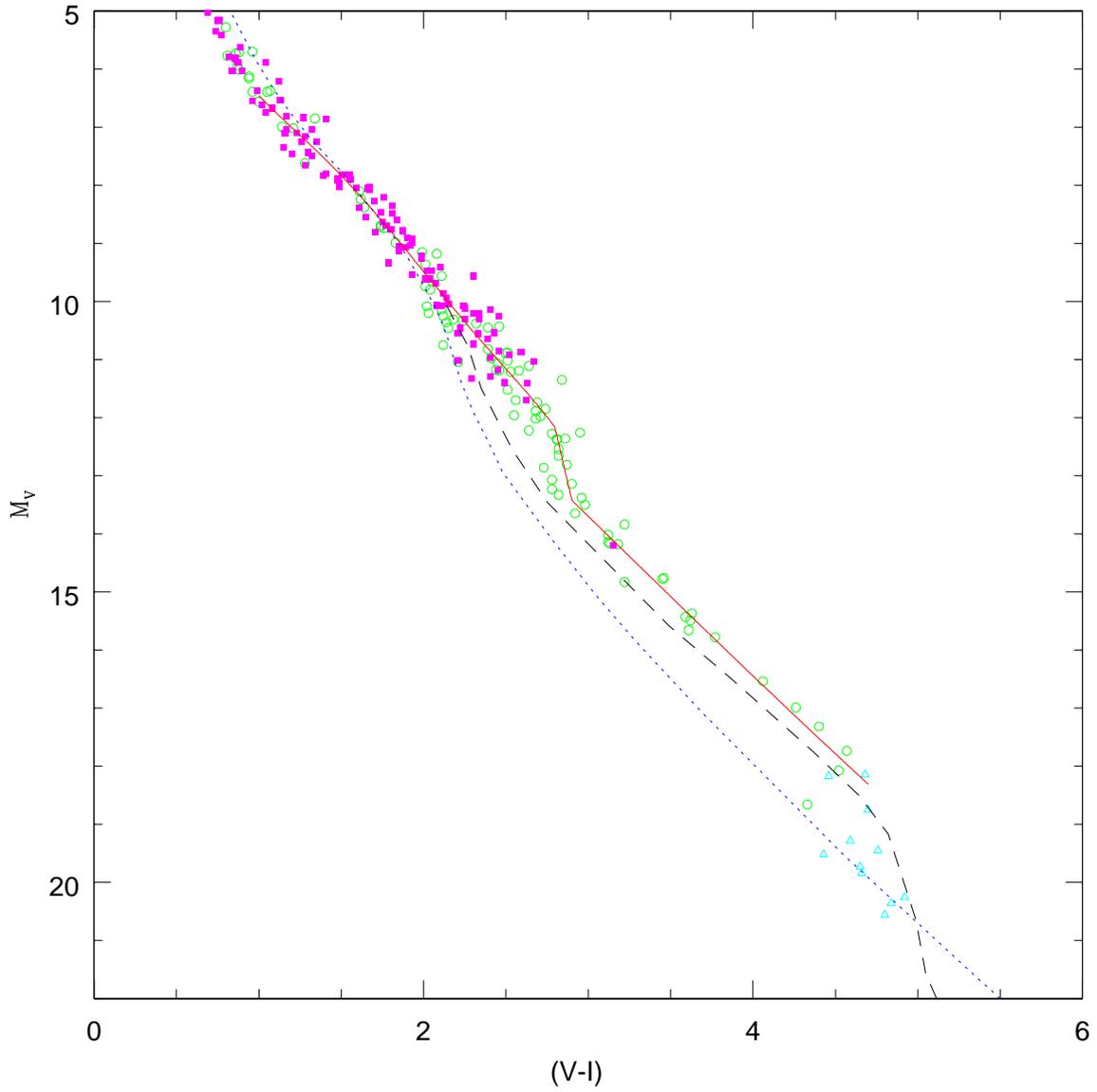}
\caption{The (M$_V$, (V-I)) relation for nearby stars:  the symbols have the same meaning as
in Figure 9, and the mean relations are given in the text
number-magnitude distribution for the 1245 stars in NLTT Sample 1. The dotted line is
the 5-Gyr. isochrone from the Baraffe {\sl et al.} (1998) models, and the dashed line
outlines the 5-Gyr. Dusty model (Chabrier {\sl et al.}, 2000). Gilles Chabrier kindly 
provided the extended isochrone, illustrating the improved agreement with observations.} 
\end{figure}

\begin{figure}
\plotone{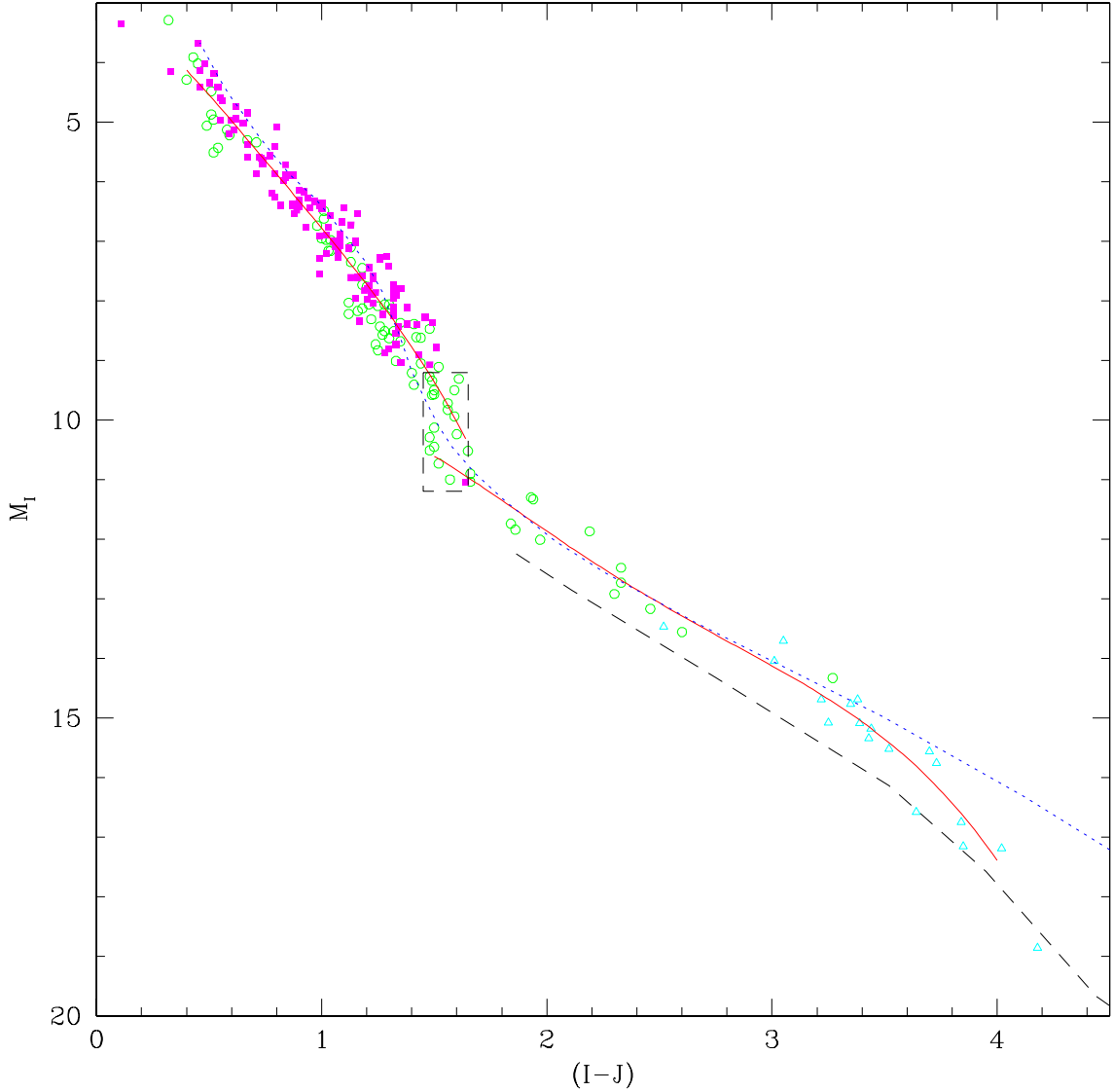}
\caption{The (M$_I$, (I-J)) relation for nearby stars: the symbols have the same meaning as in Figures
9 and 10, and the fitted relations are given in the text. We have not attempted to fit the main sequence
in the boxed region ($1.45 < (I-J) < 1.65$, $9.2 < M_I < 11.2$). The dotted line shows the 5-Gyr isochrone from
Baraffe {\sl et al.} (1998), and the dashed line plots the 5-Gyr. Dusty model  ($M < 0.1M_\odot$)
from Chabrier {\sl et al.} (2000).}
\end{figure}

\begin{figure}
\plotone{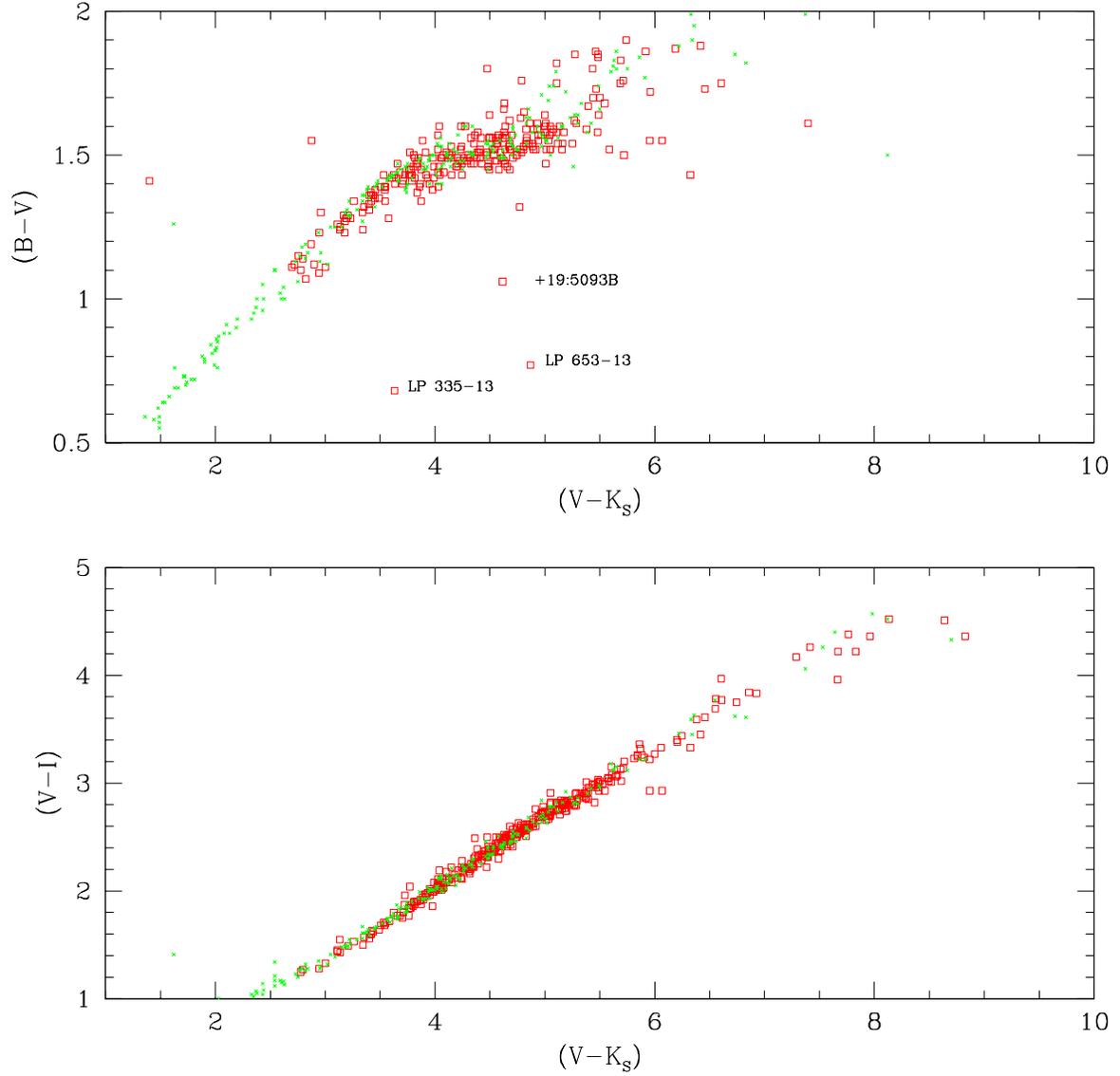}
\caption{The (B-V)/(V-K$_S$) and (V-I)/(V-K$_S$) two-colour diagrams: stars listed in
Table 2 are plotted as open squares; crosses mark the two-colour relation  defined by nearby
main-sequence stars.  The outliers are discussed in the text.}
\end{figure}

\begin{figure}
\plotone{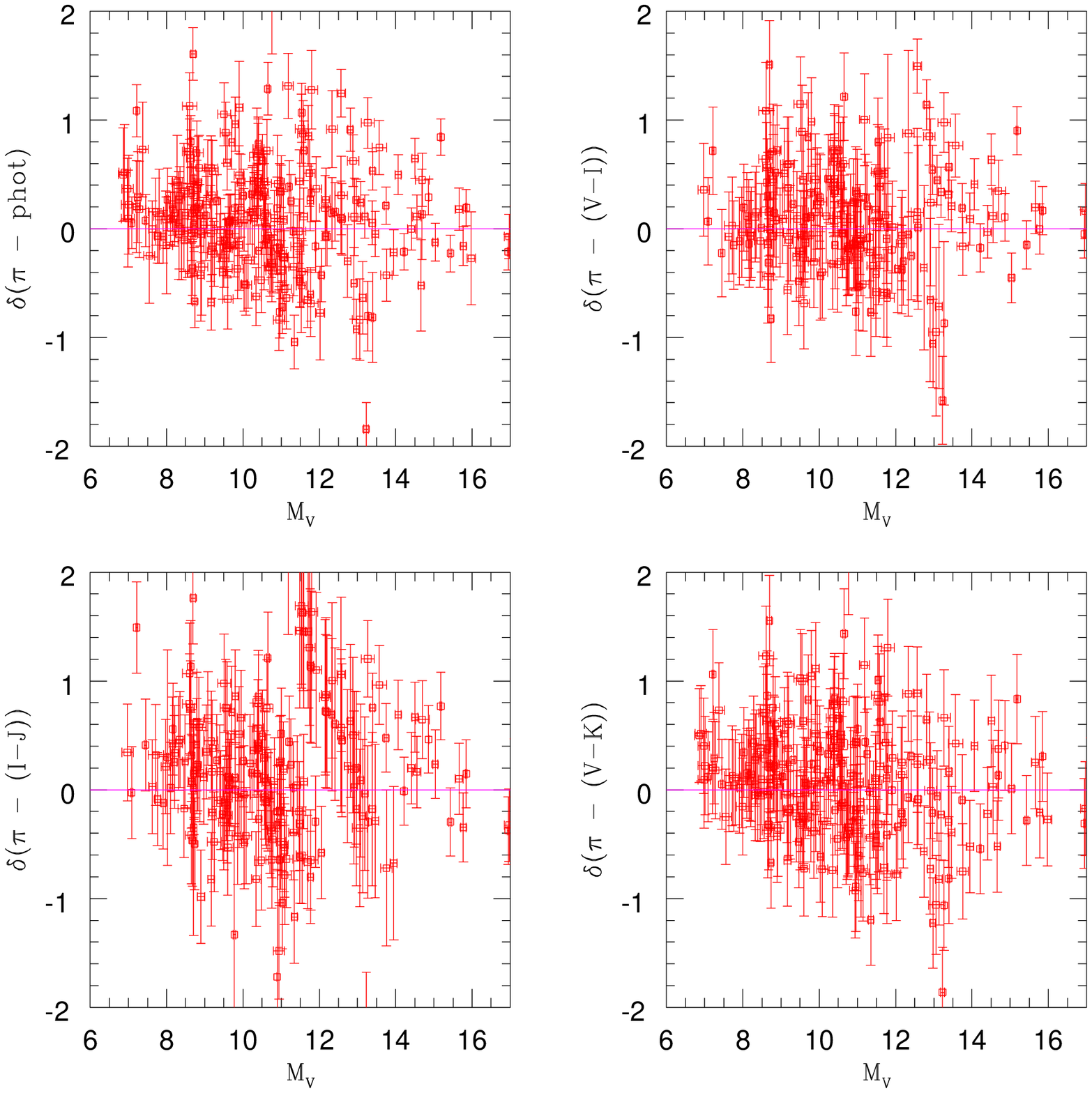}
\caption{Comparison between distance moduli derived from photometric parallaxes
and astrometric distance measurements for stars with trigonometric parallaxes measured 
to an accuracy better than 9\%. The mean residuals as a function of absolute magnitude (derived
from $\pi_{trig}$) are given in Table 4.}
\end{figure}

\newpage


\begin{references}
\overfullrule=0pt

\reference {} Andruk, V., Kharchecko, N., Schilbach, E., Scolz, R.-D., 1995, AN 316, 225

\reference {} Ardila, D., Mart\'in, E., Basri, G. 2000, AJ 120, 479

\reference {} Baraffe, I., Chabrier, G., Allard, F., Hauschildt, P. 1998, A\&A 337, 403

\reference {} Bedin, L.R., Anderson, J., King, I.R., Piotto, G. 2001, ApJL 560, L75

\reference {} Bessel, F.W. 1836, MNRAS 4, 152

\reference {} Bessell, M.S. 1990, A\&AS 83, 357

\reference {} Bessell, M.S., Brett, J.M. 1988, PASP 100, 1134

\reference {} Bessell, M.S., Weis, E.W. 1987, PASP 99, 642

\reference {} Burgasser, A.J., Kirkpatrick, J.D., Brown, M.E., Reid, I.N., Burrows, A. 
{\sl et al.} 2001, ApJ, in press

\reference {} Carpenter, J.M. 2001, AJ 121, 2851

\reference {} Chabrier, G., Baraffe, I., Allard, F., Hauschildt, P. 2000, ApJ 542, 464

\reference {} Clemens, J.C., Reid, I.N., Gizis, J.E., O'Brien, M.S. 1998, ApJ 496, 352

\reference {} Copeland, H., Jensen, J.O., Jorgensen, H.E. 1970, A\&A 5, 12

\reference {} Cruz, K.L., Reid, I.N. 2002, AJ, subm.

\reference {} D'Antona, F., Mazzitelli, I. 1985, ApJ 296, 502

\reference {} Dahn, C.C.,   Guetter, H.,  Harris, H., Henden, A., {\sl et al.} 2000, in
{\sl From Giant Planets to Cool Stars}, (ed. C. Griffiths \& M. Marley), 
ASP Conf. Proc. vol. 213, p. xxx

\reference {} Dawson, P.C. 1986, ApJ 311, 984

\reference {} Dawson, P.C., Forbes, 1989, PASP 101, 614

\reference {} Dawson, P.C., Forbes, 1992, AJ 103, 2063

\reference {} Eggen, O.J. 1966, Royal Obs. Bull. 120, 333

\reference  {} Eggen, O.J. 1975, PASP 87, 107

\reference  {} Eggen, O.J. 1980, ApJS 43, 457

\reference {} Eggen, O.J. 1987, AJ 92, 379

\reference {} Epchtein, N., De Batz, B., Copet, E. {\sl et al.} 1994, in {\sl Science with
Astronomical Near-infrared Sky Surveys}, ed. N. Epchtein, A. Omont, B. Burton, P. Persei,
(Kluwer, Dordrecht), p. 3

\reference {} ESA, 1997,  The Hipparcos Catalogue, ESA SP-1200

\reference {} Fleming, T. 1998, ApJ 504, 461

\reference {} Flynn, C., Sommer-Larsen, J., Fuchs, B., Graff, D.S., Salim, S. 2001, MNRAS 322, 553

\reference {} Geballe, T.R., Knapp, G.R., Leggett, S.K., Fan, X., Golimowski, D.A., {\sl et al.},
2001, ApJ, in press

\reference {} Giclas, H.L., Burnham, R., Thomas, N.G. 1971, The Lowell Proper Motion Survey,
Lowell Observatory, Flagstaff, Arizona

\reference {} Gizis, J.E., Monet, D.G., Reid, I.N., Kirkpatrick, J.D., Burgasser, A.J. 1999, 
MNRAS 311, 385

\reference {} Gizis, J.E., Monet, D.G., Reid, I.N., Kirkpatrick, J.D., 
Liebert, J., Williams, R.J., 2000, AJ 120, 1085

\reference {} Gliese, W. 1957, Mitt. Astron. Rechen-Inst. Heidelberg Serie A, No. 8 (CNS1)

\reference {} Gliese, W. 1969, Catalogue of Nearby Stars, Veroff. Astr. Rechen-Instituts, 
Heidelberg, Nr. 22 (CNS2)

\reference {} Gliese, W.,  Jahrei{\ss}, H. 1979, A\&AS 38, 423

\reference {} Gliese, W.,  Jahrei{\ss}, H. 1980, A\&A 85, 350

\reference {}  Gliese, W., ahrei{\ss}, H. 1991, Preliminary Version of the Third Catalogue of
Nearby Stars, (pCNS3)

\reference {} Gullixson, C.A., Boeshaar, P.C., Tyson, J.A., Seitzer, P. 1995, ApJS 99, 281

\reference {} Harrington, R.S., Fahn, C.C., Kallarakal, V.V., Guetter, H.H., Riepe, B.Y., Walker, R.L., 
Pier, J.R., Vrba, F.J., Luginbuhl, C.B., Harris, H.C., Ables, H.D. 1993, AJ 105, 1571

\reference {} Hartwick, F.D.A., Cowley, A.P., Mould, J.R.  1984,  ApJ 286, 269

\reference {} Hawley, S.L., Gizis, J.E., Reid, I.N. 1996, AJ 112, 2799 (PMSU2)

\reference {} Heintz, W.D. 1988, AJ 96, 1072

\reference {} Heintz, W.D. 1990, AJ 99, 420

\reference {} Heintz, W.D. 1991, AJ 101, 1071

\reference {} Heintz, W.D. 1993, AJ 105, 1188

\reference {} Heintz, W.D. 1994, AJ 108, 2338

\reference {} Henderson, T. 1839 MNRAS 5, 171

\reference {} Henry, T.J. 1998, in {\sl Brown dwarfs and extrasolar planets}, ASP Conf. Ser. 134
(ed. R. Rebolo, E. Mart\'in \& M. R. Zapatero Osorio), p. 28

\reference {} Humphreys, R.M., landau, R., Ghigo, F.D., Zumach, W., Labonte, A.E.  1991, AJ 102, 395

\reference {}  Kirkpatrick, J.D., Reid, I.N., Liebert, J., {\sl et al.}  1999,  ApJ 519, 802

\reference {} Kron, G.E., gascoigne, S.C.B., White, H. 1957, AJ 62, 205

\reference {} Kroupa, P., Tout, C.A., Gilmore, G. 1990, MNRAS 244, 76

\reference {} Kuiper, G.P. 1942, ApJ 95, 201

\reference {} Leggett, S.K. 1992, ApJS 82, 351

\reference {} Leggett, S.K., Allard, F., Berriman, G., Dahn, C.C., Hauschildt, P. 1996, ApJS 104, 117

\reference {} Leggett, S.K., Hauschildt, P.H., Allard, F., Geballe, T.R., Baron, E. 2002, \mnras,
in press

\reference {} Luyten, W.J. 1979, Catalogue of stars with proper motions exceeding 
0"5 annually (LHS), Univ. of Minnesota Publ., Minneapolis, Minnesota

\reference {} Luyten, W.J., Albers, H. 1979, The LHS Atlas, Univ. of
Minnesota Publ., Minneapolis, Minnesota

\reference {} Luyten, W.J. 1980, Catalogue of stars with proper motions exceeding 
0"2 annually (NLTT), Univ. of Minnesota Publ., Minneapolis, Minnesota

\reference {}  Mart\'in, E.L., Delfosse, X., Basri, G., Goldman, N.,
{\sl et al.} 1999, ApJ, 118, 2466

\reference {} Minkowski, R., Abell, G.O. 1963, in {\sl Stars and Stellar Systems, Vol. 3},
Basic Astronomical Data, ed. K.Aa. Strand, (Chicago, Univ. of Chicago Press), p. 481

\reference {} Monet, D.G., Dahn, C.C., Vrba, F.J., Harris, H.C., Pier, J.R., Luginbuhl, C.B., 
Ables, H.D. 1992, AJ 103, 638

\reference {} Patterson, R.J., Ianna, P.A., Begam, M.C. 1998, AJ 115, 1648

\reference {} Persson, S.E., Murphy, D.C., Krzeminski, W., Roth, M., Rieke, M.J. 1998, AJ 116, 2475

\reference {} Reid, I.N. 1990, MNRAS 247, 70

\reference {} Reid, I.N., Gizis, J.E. 1997, AJ 113, 2246

\reference {} Reid, I.N., Hawley, S.L., Gizis, J.E. 1995, AJ 110, 1838 (PMSU1)

\reference {} Reid, I.N., Sahu, K.C., Hawley, S.L. 2001, ApJL, in press

\reference {} Reid, I.N., Cruz, K.L. 2001, AJ, subm.

\reference {} Reid, I.N., Kilkenny, D., Cruz, K.L. 2001, AJ subm.

\reference {} Ryan, S.G. 1989, AJ 98, 1693

\reference {} Ryan, S.G. 1992, AJ 104, 1144

\reference {} Sandage, A, Kowal, C. 1986, AJ 91, 1140

\reference {} Scholz, R.-D., Meusinger, H., Jahrei{\ss}, H. 2001, A\&A, in press

\reference {} Skrutskie, M.F. {\sl et al.} 1997, in {\sl The Impact of Large-Scale
Near-IR Sky Survey}, ed. F. Garzon et al (Kluwer:  Dordrecht), p. 187

\reference {} Stauffer, J., Hartmann, L. 1986, ApJS 61, 531

\reference {} Tinney, C.G. 1996, MNRAS 281, 644

\reference {} Tinney, C.G. 1998, MNRAS 296, L42

\reference {} Upgren, A.R., Lu, P.K. 1986, AJ 92, 903

\reference {} van Altena, W.F., Lee, J.T., Hoffleit, E.D. 1995, Yale Catalogue of
Trigonometric Parallaxes, 4th edition, (Yale University Observatory)

\reference {} van de Kamp, P. 1940, Pop. Astr. 48, 297

\reference {} Weis, E.W. 1984, ApJS 55, 289

\reference {} Weis, E.W. 1986, AJ 91, 626

\reference {} Weis, E.W. 1987, AJ 92, 451

\reference {} Weis, E.W. 1988, AJ 96, 1710

\reference {} Weis, E.W. 1991, AJ 102, 1795

\reference {} Weis, E.W. 1993, AJ 105, 1962

\reference {} Weis, E.W. 1996, AJ 112, 2300

\reference {} Weis, E.W. 1999, AJ 117, 3021

\end{references}
\end{document}